\documentclass[1p]{elsarticle}
\makeatletter
\def\ps@pprintTitle{%
 \let\@oddhead\@empty
 \let\@evenhead\@empty
 \def\@oddfoot{\centerline{\thepage}}%
 \let\@evenfoot\@oddfoot}
\makeatother
\usepackage{lineno}
\usepackage{comment}
\usepackage{amsmath}
\usepackage{graphicx}
\usepackage{amsfonts}
\usepackage[colorinlistoftodos]{todonotes}
\usepackage[colorlinks=true, allcolors=blue]{hyperref}
\usepackage{subcaption}
\usepackage{booktabs}
\usepackage{tikz}
\usepackage{makecell}
\usepackage{multirow}
\usepackage{geometry}
\usepackage{mathtools}
\usepackage[ruled,vlined]{algorithm2e}

\usetikzlibrary{shapes.geometric, arrows}
\usetikzlibrary{snakes,trees}
\usetikzlibrary{mindmap}
\usetikzlibrary{calc}
\usetikzlibrary{shapes,arrows,decorations.pathmorphing,positioning,fit}

\usetikzlibrary{calc,trees,positioning,arrows,chains,shapes.geometric,%
 decorations.pathreplacing,decorations.pathmorphing,shapes,%
    matrix,shapes.symbols,shapes.geometric,backgrounds,calc}
    
\tikzstyle{subprocess} = [rectangle, minimum width=1cm, minimum height=0.27cm, text centered, draw=black] 
\tikzstyle{state} = [rectangle, minimum width=1cm, minimum height=0.27cm, text centered, draw=black]
\tikzstyle{terminal} = [rectangle,rounded corners, minimum width=1cm, minimum height=0.27cm, text centered, draw=black]
\tikzstyle{arrow} = [thick,->,>=stealth]
\tikzstyle{dasharrow} = [thick,dashed,->,>=stealth]
\tikzstyle{dasharrownohead} = [thick,dashed,-,>=stealth]
\tikzstyle{textbf} = [draw,rectangle,text width=5cm,text centered]

\tikzset{
  basic box/.style = {
    shape = rectangle,
    align = center,
    draw  = #1,
    fill  = #1!25,
    rounded corners},
  header node/.style = {
    Minimum Width = header nodes,
    font          = \strut\Large\ttfamily,
    text depth    = +0pt,
    fill          = white,
    draw},
  header/.style = {%
    inner ysep = +1.5em,
    append after command = {
      \pgfextra{\let\TikZlastnode\tikzlastnode}
      node [header node] (header-\TikZlastnode) at (\TikZlastnode.north) {#1}
      node [span = (\TikZlastnode)(header-\TikZlastnode)]
        at (fit bounding box) (h-\TikZlastnode) {}
    }
  },
  hv/.style = {to path = {-|(\tikztotarget)\tikztonodes}},
  vh/.style = {to path = {|-(\tikztotarget)\tikztonodes}},
  fat blue line/.style = {ultra thick, blue}
}    

\biboptions{authoryear}

\begin{document}

\begin{frontmatter}

\title{Evaluating congestion pricing schemes using agent-based passenger and freight microsimulation}


\author[mymainaddress]{Peiyu Jing}
\author[mythirdaddress]{Ravi Seshadri\corref{mycorrespondingauthor}}
\cortext[mycorrespondingauthor]{Corresponding author}
\ead{ravse@dtu.dk}
\author[myfourthaddress]{Takanori Sakai}
\author[mymainaddress]{Ali Shamshiripour}
\author[mymainaddress]{Andre Romano Alho}
\author[mymainaddress]{Antonios Lentzakis}
\author[mymainaddress]{Moshe E. Ben-Akiva}

\address[mymainaddress]{Civil and Environmental Engineering Department, Massachusetts Institute of Technology, Cambridge, MA, US}
\address[mythirdaddress]{Department of Technology, Management and Economics, Technical University of Denmark, Denmark}

\address[myfourthaddress]{Department of Logistics \& Information Engineering,
Tokyo University of Marine Science and Technology, Japan}


\begin{abstract} 
The distributional impacts of congestion pricing have been widely studied in the literature and the evidence on this is mixed. Some studies find that pricing is regressive whereas others suggest that it can be progressive or neutral depending on the specific spatial characteristics of the urban region, existing activity and travel patterns, and the design of the pricing scheme. Moreover, the welfare and distributional impacts of pricing have largely been studied in the context of passenger travel whereas freight has received relatively less attention. In this paper, we examine the impacts of several third-best congestion pricing schemes on both passenger transport and freight in an integrated manner using a large-scale microsimulator (SimMobility) that explicitly simulates the behavioral decisions of the entire population of individuals and business establishments, dynamic multimodal network performance, and their interactions. Through simulations of a prototypical North American city, we find that a distance-based pricing scheme yields the largest welfare gains, although the gains are a modest fraction of toll revenues (around 30\%). In the absence of revenue recycling or redistribution, distance-based and cordon-based schemes are found to be particularly regressive. On average, lower income individuals lose as a result of the scheme, whereas higher income individuals gain. A similar trend is observed in the context of shippers -- small establishments having lower shipment values lose on average whereas larger establishments with higher shipment values gain. We perform a detailed spatial analysis of distributional outcomes, and examine the impacts on network performance, activity generation, mode and departure time choices, and logistics operations.

\end{abstract}


\end{frontmatter}


\section{Introduction}

In 2018, transportation accounted for around 24\% of global energy-related carbon dioxide emissions, with road transportation contributing three-quarters of this amount\footnote{https://ourworldindata.org/co2-emissions-from-transport}, and in the United States in 2020, transportation accounted for 27\% of anthropogenic greenhouse gas emissions\footnote{https://www.epa.gov/greenvehicles/fast-facts-transportation-greenhouse-gas-emissions}.  
Currently, 50\% of the world’s population resides in urban areas, and the rapid rate of urbanization on a global scale has led to an increased demand for urban transportation. This proportion is expected to grow to 70\% by 2050\footnote{United Nations (2018). World Urbanization Prospects. https://population.un.org/wup/.}, and consequently, competition for road surface area will likely remain stiff in the ensuing decades.
Cities are now, more than ever, contending with growing congestion and its consequences. In 2014 alone, U.S. auto commuters together lost an estimated 6.9 billion hours and 3.1 billion gallons of fuel to congestion—42 hours and 19 gallons wasted per commuter \citep{schrank20152015}. The global urban average in 2017 was not much better: 27 hours and 13 gallons per driver \citep{cookson2017inrix}. The rapid worldwide growth of on-demand mobility has also contributed to congestion. In the U.S., on-demand mobility has grown steadily, and its rise has led to the cannibalization of mass transit ridership by 6\% from 2014 through 2016 \citep{circella2018transport}. Similar downward trends in mass transit ridership have been observed on a global scale\footnote{UITP, (2017). Urban Public Transport in the 21st Century: Statistics Brief. https://bit.ly/2SihK9A.}. E-commerce has also grown in significance over the past decade. A greater number of deliveries are made to consumers, driving the demand for urban freight. In addition to competing for existing road capacity, carriers make multiple stops along their routes, further exacerbating congestion and pollution.

The growing consensus is that demand management solutions are the most viable strategies to mitigate congestion whereas supply-oriented solutions (typically driven by road capacity expansion) are often less attractive and feasible, owing to their high costs and low impact \citep{gu2018congestion}. In particular, congestion pricing schemes --the standard economic prescription to internalize costs of negative externalities-- are advantageous in that they encourage travelers to adjust all facets of their behavior including trip making, mode, departure-time and destination choices, as well as residential and work location \citep{de2011traffic}. Notable examples of congestion pricing schemes in practice include the cities of Singapore, Stockholm, Gothenburg, Milan and London (termed Downtown Congestion pricing by \cite{lehe2019downtown}), and High-Occupancy/Express/Managed Lane facilities in the United States. Broadly, these schemes can be classified into four categories: facility-based, cordon, zonal, and distance-based schemes (see \cite{de2011traffic} for a more detailed taxonomy). Extensive reviews of congestion pricing may be found in   \cite{de2011traffic,santos2011road,anas2020reducing}.   

The welfare and distributional effects of congestion pricing (more generally, road pricing) have been widely studied in the literature. While it is generally the case that congestion pricing will generate a net welfare gain, the efficiency gains can often be dwarfed by distributional impacts \citep{eliasson2006equity,arnott1994welfare}. The regressivity of pricing is often the focal point of public and political resistance to congestion pricing  (for a detailed discussion on the political calculus underlying congestion pricing we refer the reader to \cite{king2007political} and for more on equity and the different notions of progressivity, see \cite{arnott1994welfare} and \cite{levinson2010equity}). Nevertheless, the consensus from the literature on this is mixed. Several researchers have argued that congestion pricing is regressive in that prior to a redistribution of revenues, individuals with a high income and a higher value of time fare better than those with lower values \citep{evans1992road,arnott1994welfare,seshadri2022congestion}. Individuals with smaller economic margins generally have less flexibility in avoiding the toll charges (for example, by changing departure time), are more likely to live far from the city center and have destinations in areas where public transport is relatively poor \citep{eliasson2006equity}. In contrast, and often in the context of European cities with better developed transit systems, researchers have found that pricing can be progressive \citep{eliasson2006equity,santos2004distributional}. Ultimately, the question of distributional impacts of pricing is likely to depend on the spatial structure of the city in question, socioeconomic characteristics and differences, and their relationship with activity and travel patterns, the design of the charging scheme, and so on (see \cite{eliasson2006equity,santos2004distributional}). Moreover, assessing the performance of pricing schemes in general networks is complicated by network topology and the interdependence of flows, and discrepancies in findings from simple versus complex models in the literature point to the difficulty in drawing general conclusions \citep{de2011traffic}. 

The aforementioned considerations all motivate the need for evaluating the performance of congestion pricing schemes using large-scale disaggregate microsimulation models. Further, the increasing role of freight and e-commerce in urban congestion underscores the importance of examining the impacts on both passenger transport and freight. In this regard, this paper contributes to the literature in the following respects:  

\begin{itemize}
\item We propose an approach for the synthesis of business establishments for freight microsimulation models using publicly available data and apply the approach to generate a population of synthetic business establishments for Boston
\item We comprehensively evaluate the impact of three \textit{third-best} congestion pricing schemes (area, distance, cordon) using an integrated passenger and freight agent-based microsimulation model of a large-scale prototype North American city (involving a population of 4.6 million individuals and 0.130 million business establishments)
\item We analyze welfare and distributional impacts on passengers and shippers, network performance impacts, impacts on activity patterns, and the effect on logistics operations 
\end{itemize}

To the best of our knowledge, our study is the first to examine the impacts of congestion pricing on both passenger and freight in an integrated manner using a large-scale microsimulation model that explicitly simulates the behavioral decisions of the entire population of individuals and business establishments, dynamic multimodal network performance, and their interactions.

\section{Review of Literature}\label{sec:Sec2}
The literature on congestion pricing is vast and includes the design and assessment of pricing using dynamic bottleneck models \citep{arnott1990economics,arnott1994welfare,van2011congestion,van2011winning,verhoef1996second,verhoef2004product}, design of first-best pricing in general networks \citep{yang1998principle,yang1999system}, design of second-best pricing schemes in general networks \citep{verhoef2002second,zhang2004optimal}, empirical assessments of congestion pricing and evaluation of pricing used large-scale static and dynamic network models \citep{eliasson2006equity,eliasson2016congestion,de2005congestion,santos2004distributional}. We restrict our attention to two streams of research, namely the study of congestion pricing using large-scale dynamic simulators and the impacts of pricing on freight. We refer the reader to the \cite{santos2011road, de2011traffic} for more detailed reviews.   

\subsection{Impacts of pricing using large-scale dynamic simulators}
There exist several large-scale agent-based simulation models that have been developed over the years. These include MATSim \citep{w2016multi}, POLARIS \citep{auld2016polaris}, SimMobility \citep{oh2021impacts,adnan2016simmobility}, METROPOLIS \citep{de1997metropolis}, etc. However, detailed quantitative assessments of congestion pricing using such models is surprisingly sparse. 

\cite{de2005congestion} analyzes several link-based road pricing schemes using the dynamic network simulator METROPOLIS, which endogenously models departure time, route and mode choice decisions. Using a stylized urban road network consisting of radial arterials and circumferential ring roads, they find that step tolls outperform flat tolls and generate smaller revenues, leading to more favourable distributional outcomes. 

Several studies have used the multi-agent simulation platform MATSim to examine the impacts of congestion pricing schemes in different contexts. \cite{kaddoura2019congestion} examines the impacts of two pricing schemes, one utilizes the marginal external congestion costs based on the queuing dynamics at bottleneck links and the second uses a control-theoretical approach to adjust toll levels depending on the congestion level. They perform experiments using a case study of the Greater Berlin region by simulating 10\% of the total population with appropriately reduced link capacities. The findings suggest that the pricing policies cannot completely eliminate queuing and that the overall welfare gains are between 2.5\% and 14\% of toll revenues. \cite{simoni2019congestion} examines the impacts of several congestion pricing schemes for future scenarios with automated and shared automated vehicles. They employ agent-based simulations of the city of Austin and its surroundings (once again, as in   \cite{kaddoura2019congestion}, they simulate a fraction of the total population, in this case 5\%), and find that a travel time-congestion-based scheme yields the highest welfare gains. Finally, more recently, \cite{he2021validated} study the impacts of pricing schemes in New York City using a calibrated NYC MATSim model. They use a 4\% sample of the population (around 320000) for simulation. In contrast with most previous studies, they find a net increase in consumer surplus prior to any redistribution of toll revenues and attribute this to heterogeneity in values of time. 

\cite{lentzakis2020hierarchical} and \cite{lentzakis2023predictive} demonstrate the superiority of distance-based congestion pricing schemes (in terms of overall welfare gains) over area- and cordon-based schemes using simulations of the network of Singapore expressways and downtown Boston. They employ the real-time Dynamic Traffic Assignment model system DynaMIT2.0 \citep{zhang2017improved,zhang2021improving} considering departure time and route choices under elastic demand.

\subsection{Impacts of pricing on Freight}

The impact of pricing on freight demand has not been studied as much as that on passenger demand. \cite{waliszewski2005towards} surveyed the potential impacts of congestion pricing on urban freight based on available data. Although London’s congestion charging scheme resulted in a reduction in the number of trucks, the proportion of trucks among all road traffic increased. This suggests a lower price elasticity of freight demand relative to passenger demand. The cases in New York City, Switzerland, and Austria suggest that tolling trucks for externalities would lead to more equitable and efficient truck operations. In Switzerland and many European cities where distance-based charging is implemented for trucks, evidence suggests that long-distance travel becomes less efficient and frequent. \cite{holguin2006impacts} assessed a time-of-day pricing initiative on commercial carriers for toll facilities (i.e., bridges and tunnels) operated by the Port Authority of New York and New Jersey. The authors note that the carrier responses to the pricing are complex and heterogeneous, involving changes in facility usage (changes in the amount of toll facility usage), productivity increases (behavioral changes that increase efficiency and productivity), and cost transfers to the end consumer. They indicated that the constraints that prevent carriers from changing operations from peak-hours to non-peak-hours are significant. \cite{holguin2010truth} corroborated this by analyzing interactions between shippers and carriers with empirical and game theoretical evidence. He concluded that freight mode choice and delivery time choice are jointly made by shippers and carriers, so carrier-centered pricing schemes may not always achieve the desired impacts. Furthermore, \cite{holguin2011urban} modeled the cost transfer between carriers and receivers and derived theoretical impacts of cordon pricing, time–distance pricing, and carrier–receiver policies on the urban delivery industry in competitive markets. The results suggest that cordon time-of-day pricing cannot transfer costs to receivers and is thus ineffective, whereas time-distance pricing could transfer costs to receivers with huge unit tolls.

To quantitatively evaluate the impact of urban congestion pricing on freight and passenger traffic together, \cite{chen2018impact} developed a transportation network equilibrium model and quantified the temporal shift of freight traffic under congestion pricing. They use a network that is relatively simple with two origins and one destination. The results highlight that, with increasing VOT, the equilibrium flows decrease and the average speed increases, indicating better effectiveness in congestion reduction and network throughput. Although VOT is the key consideration in decision-making and measuring cost sensitivity, \cite{chen2018impact} assume a single VOT for each simulation. Finally, to understand the heterogeneity among drivers, \cite{toledo2020intercity} develop a mixed logit model with a path-size factor for truck driver route choices, incorporating heterogeneity by employment term and vehicle characteristics.

In summary, to the best of our knowledge, there are no integrated quantitative assessments of congestion pricing on both passenger transport and freight, the impacts on freight have been less studied, and moreover, applications of large-scale agent-based simulators have tended to use a small fraction of the population for simulation, which may or may not accurately capture system-wide impacts.  

\begin{figure}[!ht]
\centering
\hspace{-1 cm}
\includegraphics[scale=0.4]{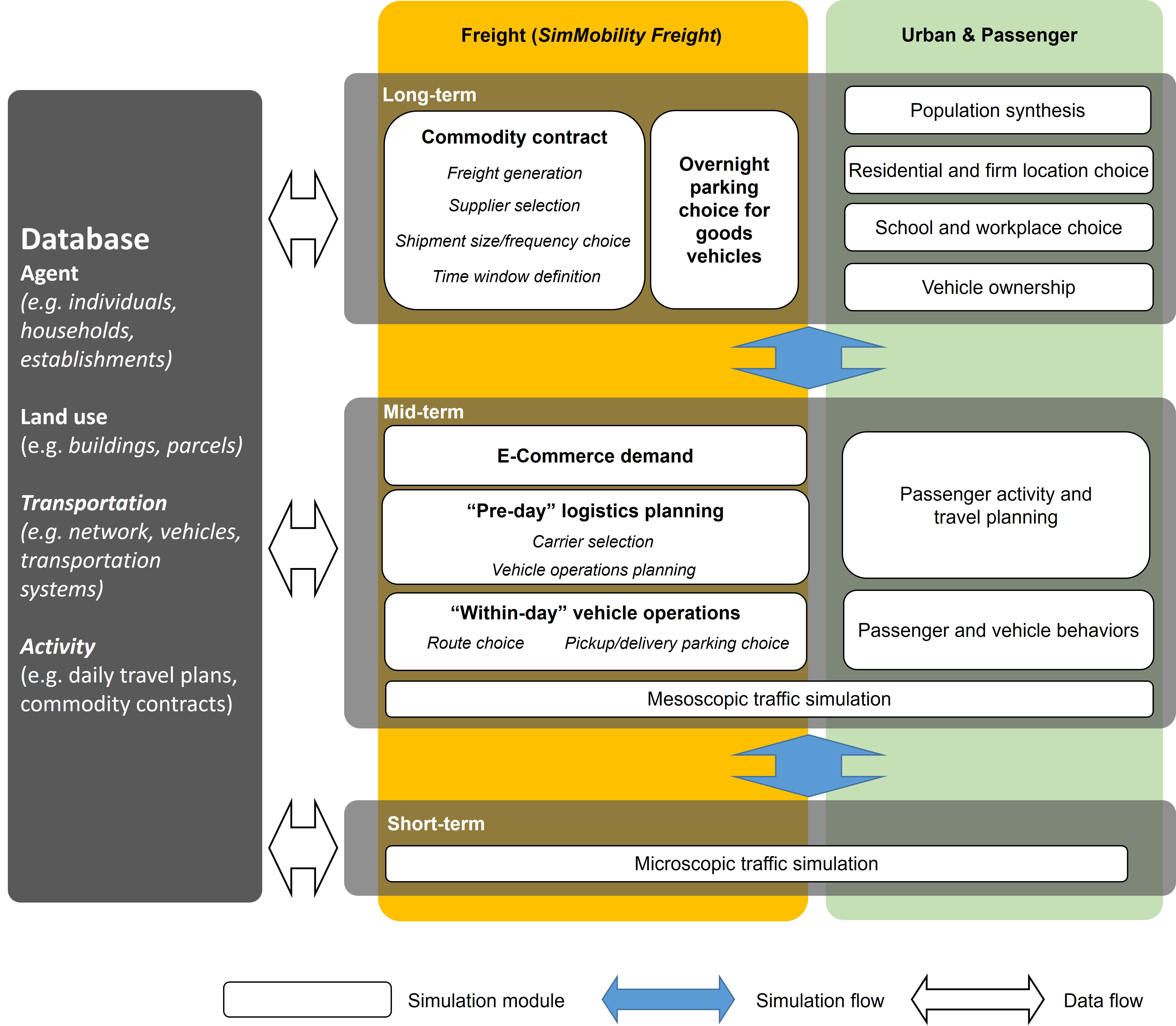}
  \caption{SimMobility Framework }  \label{fig:SM_Framework}
\end{figure}  

\section{Agent-based Microsimulation Model}\label{sec:Sec3}
In assessing the impacts of congestion pricing schemes, it is imperative that in addition to overall welfare gains, distributional impacts are reliably quantified and `winners´ and `losers´ from the policy are identified. Equity issues associated with congestion pricing have long been recognized \citep{levinson2010equity} and mitigating the regressive effects of pricing remain the key to ensuring public acceptability. This motivates the need for a disaggregate microsimulation-based approach that explicitly models the behavior of individual passenger and freight agents. For this, we utilize an open-source agent-based passenger and freight microsimulator, SimMobility \citep{adnan2016simmobility}. In this section, we provide a concise description of SimMobility and detail enhancements on the freight side to model price sensitivity. 

\subsection{Overview}
SimMobility simulates agents’ behavioral decisions at three different temporal scales - long-term, mid-term, and short-term (Figure \ref{fig:SM_Framework}). The long-term model simulates strategic decisions involving time periods longer than a day (e.g., residential locations, business locations, and commodity contracts). 
On the passenger side, agents are individuals and households whereas on the freight side, each establishment is an agent, and can play the role of a shipper, a carrier, a receiver, or a combination of these. The long-term model includes population synthesis, the simulation of residential, school, workplace, business location choice, and vehicle ownership for passengers and business establishments. A business establishment also decides on commodity contracts (that determine business-to-business (B2B) commodity flows between establishments) and chooses overnight parking locations of freight vehicles. Here, a commodity contract defines the origin and destination of a commodity flow, commodity type, contract size, shipment size and frequency, etc. 

The mid-term model simulates individuals' activities and freight deliveries at the day level. The mid-term model combines activity-travel based demand models with multimodal dynamic, mesoscopic network assignment. The demand simulator comprises two groups of behavior models: pre-day and within-day. The \textit{pre-day} models simulate the plans of all the individuals and freight-related establishments, as well as e-commerce orders, while within-day models adapt these plans based on the current network conditions and constraints. The \textit{Supply} module then simulates the movement of individuals and vehicles (mass transit, on-demand, and freight) and their interactions based on a combination of macroscopic speed-density relationships and queueing models. The decisions of supply agents such as public transit and on-demand service operators are explicitly modeled (\cite{oh2020assessing}).

Finally, the short-term model is a microsimulator of transportation operations with high temporal resolution including models of car-following and lane-changing. The simulation time step is typically a fraction of a second. 

\begin{figure}[!ht]
\centering
\hspace{-1 cm}
\includegraphics[scale=0.8]{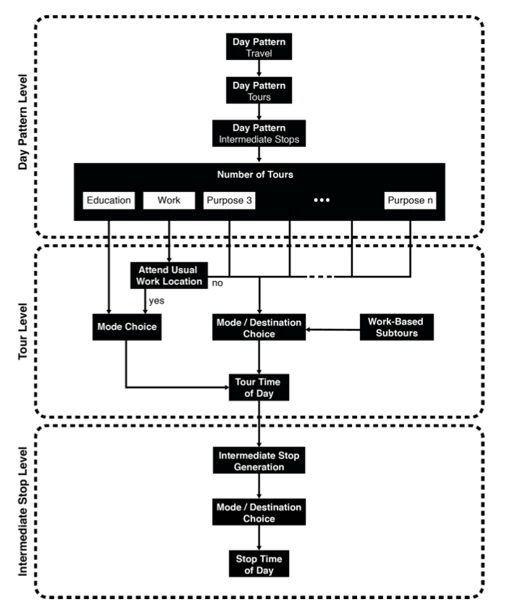}
  \caption{Pre-day Activity-based Model System}  \label{fig:SM_MT_Preday}
\end{figure}
\subsection{Passenger Demand}
The pre-day passenger demand module is an activity-based model system that adopts the day-activity schedule approach \citep{bowman2001activity}. 
It predicts an activity-travel schedule (plan) for each individual in the population including the number and types of activities performed, their duration and time-of-day,
locations, and travel modes. As shown in Figure \ref{fig:SM_MT_Preday}, it is a system of hierarchical discrete choice models (logit and nested-logit) organized into three levels: the day-pattern level, the tour level and the intermediate stop level. Bottom-level
decisions are conditional on upper-level decisions and the levels are related through ‘logsums' or ‘inclusive values’ (expected maximum utility) indicated by dashed lines. The inclusive values capture the sensitivity of activity and travel decisions modeled at the higher levels of the hierarchy to the utility associated with conditional outcomes from the lower level models. All the stop and tour-level models are sensitive to pricing and consequently, so are the upper-level day pattern decisions through the logsums. More details on the specification and calibration of the models are provided in Section \ref{sec:Sec4}.

The within-day module transforms the pre-day plans into actions and includes path-size logit models of private and public route choice. The attributes in the private route choice model includes travel time, toll cost, distance, number of signalized intersections, and number of right turns. The attributes in the public-transit route choice model includes in-vehicle time, wait-time, walk-time, and number of transfers.  We refer the reader to \cite{adnan2016simmobility} and \cite{tan2015new} for more details on the structure of the route choice models.

\subsection{Freight Demand}
\begin{figure}[!ht]
\centering
\hspace{-1 cm}
\includegraphics[scale=0.42]{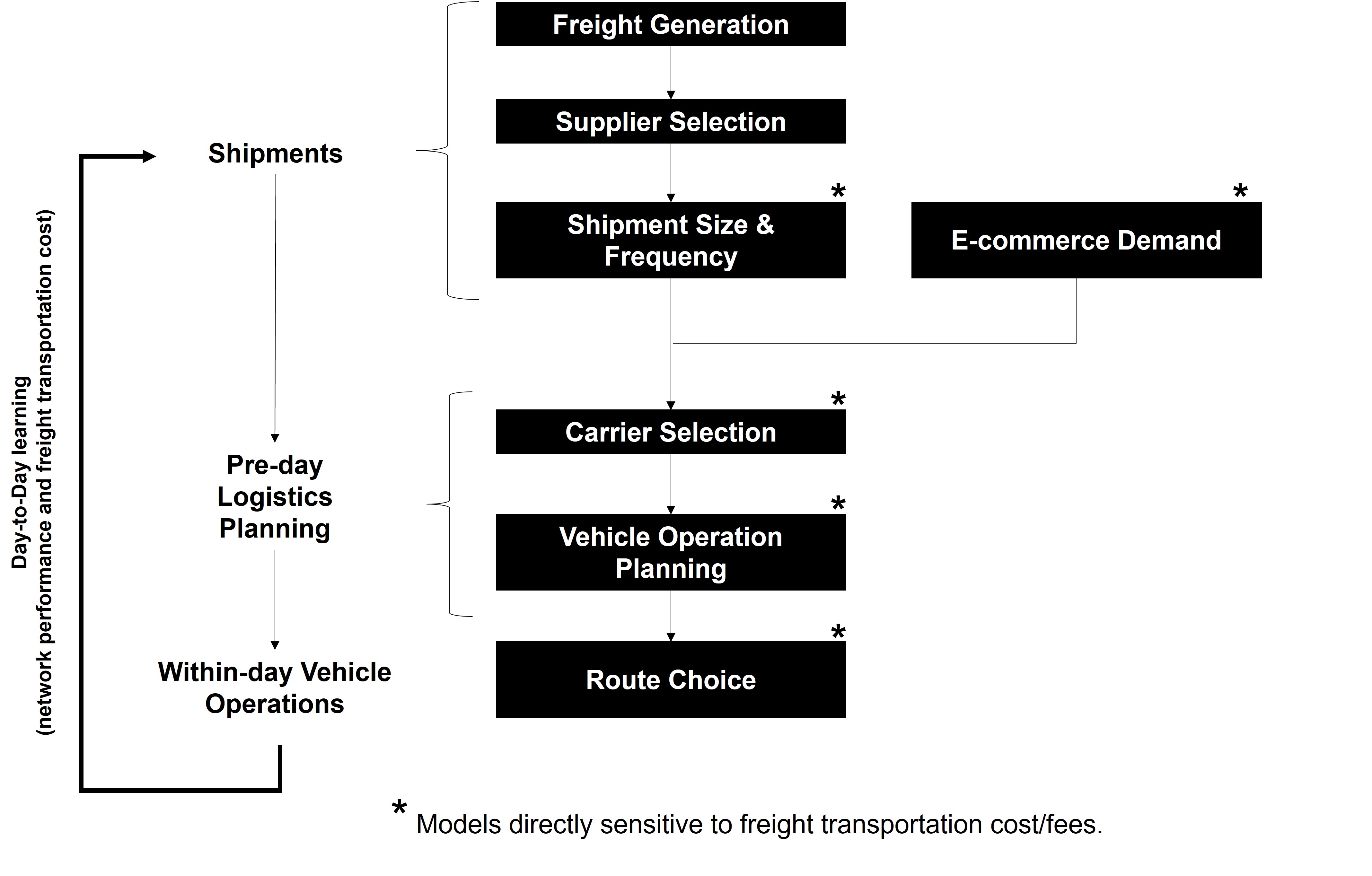}
  \caption{Price sensitive models in SimMobility Freight}  \label{fig:sm_freight_sensitive}
\end{figure}

A detailed description of the structure of the freight demand models may be found in \cite{sakai2020simmobility}, while the details of the household-based e-commerce model are available in \cite{sakai2022household}. In the long-term, establishments (either as receivers or shippers or both ) generate demands for B2B commodity flows through freight generation (production and consumption), supplier selection, and shipment size and frequency selection. At the mid-term pre-day level, households and e-commerce vendors (as shippers) generate B2C shipments, shipper establishments select carrier establishments, and finally, carriers plan their vehicle operations. The process of assigning shipments to vehicles and developing vehicle tours is called vehicle operations planning (VOP). At the mid-term within-day level, freight vehicle drivers make route and parking choices considering real-time traffic information. In the decision-making of establishments (and their drivers) and households, we assume they are independent entities that make decisions individually. Admittedly, this is a simplistic assumption and one we make due to limited knowledge about their interactions. The key models (at the long-term and mid-term level) sensitive to congestion pricing are summarized in Figure \ref{fig:sm_freight_sensitive}. We assume that overall freight generation is not affected considering that city-level pricing schemes have negligible impact on regional or global B2B commodity production and consumption, whereas shippers’ shipment size and frequency decisions, and e-commerce demand are affected. Carriers are also sensitive to costs in VOP. In the within-day level, the drivers’ route choice is sensitive to toll costs. Next, we describe enhancements to the models in order to incorporate price sensitivity. Note that in this section, we primarily describe model structure. Details on the estimation and calibration of model parameters are provided in Section \ref{sec:Sec4}. 

\subsubsection{Shipment size and Frequency}
The shipment size and frequency model is an enhanced version of the model in \cite{sakai2020simmobility}. It determines the shipment sizes in commodity contracts, and given the annual commodity demand, also generates the frequency – the number of shipments annually and on an average weekday. The model relies on economic order quantity theory. A shipper $s$ optimizes the shipment size to minimize the annual logistics cost of a contract, where three cost components are relevant to the quantity. 
The total logistics cost ($TLC$) for a contract $i$ is defined as:
\begin{equation}
TLC_i=T_i+I_i+K_i,
\end{equation}
where $I_i$ is the inventory cost incurred by receiver $r$, $K_i$ is the capital cost during transportation, and $T_i$ is the transportation cost incurred by the receiver. It should be noted that other components such as cost of deterioration and damage during transport, capital cost during transport, and stockout cost, are not included because they are not functions of shipment size and do not affect the shipment size selection. For simplicity, the subscript $i$ is omitted in following discussion. The inventory cost is assumed to be:
\begin{equation}
I = w_{s,r}^{com}\times q/2,
\end{equation}
where $w_{s,r}^{com}$ is the commodity type and location specific inventory cost per weight unit, and $q$ is the shipment size by weight. $w_{s,r}^{com}$ is assumed to be linear in the establishment density at the receiver’s location:
\begin{equation}\label{eqn:WET}
w_{s,r}^{com} = \beta_{ED}^{com}\times ED_r,
\end{equation}
where $ED_r$ is the establishment density at the receiver location and  $\beta_{ED}^{com}$ is a unknown parameter. The capital cost is:
\begin{equation}
K = d\times v^{com}\times q/2,
\end{equation}
where $d$ is the discount rate, $v^{com}$ is the commodity-type-specific goods value per unit of weight. Finally, transport cost is given by:
\begin{equation}
T=(Q/q)\times o_{s,r}^{com},
\end{equation}
where $Q$ is the annual commodity flow by weight, $o_{s,r}^{com}$ is the commodity type and OD specific transport cost. We define $o_{s,r}^{com}$ as a function of average operation costs (including tolls):
\begin{equation}
o_{s,r}^{com} = \left( \beta_{q_0}^{com}+\beta_q^{com}q \right) \times AOC_{s,r}^{com},
\end{equation}
where $AOC_{s,r}^{com}$ is calculated as the total operation cost divided by the total number of shipments for each combination of OD and commodity type, and is updated iteratively within the day-to-day learning process described in Section \ref{sec:Supply}. $\beta_{q_0}^{com}$ and $\beta_q^{com}$ are unknown parameters.  The detailed definition of the operation cost is discussed in Section \ref{sec:Sec5}. From the above equations, the optimal shipment size is given by:
\begin{equation}
q = \left( \frac{2\times \beta_{q_0}^{com}\times Q \times AOC_{s,r}^{com} }{ \beta_{ED}^{com}\times ED_r +  d\times v^{com}  } \right) ^{0.5},
\end{equation}
where  $\beta_{q_0}^{com}$ and $\beta_{ED}^{com}$  are model parameters. With congestion pricing, if $AOC_{s,r}^{com}$ increases, then the shipment size increases. It is expected that the size change would be more significant for contracts that have lower commodity value and for ODs with higher cost.
It should be noted that since the calibration of only $\beta_{q_0}^{com}$ and $\beta_{ED}^{com}$ in equation 7 cannot reproduce the observed shipment size distribution (this issue is pointed also by \cite{sakai2020empirical} using the data from Tokyo, Japan), we introduce an additional parameter $\beta_Q$:

\begin{equation}\label{eqn:SMS}
q = \left( \frac{2\times \beta_{q_0}^{com}\times Q^{\beta_Q} \times AOC_{s,r}^{com} }{ \beta_{ED}^{com}\times ED_r +  d\times v^{com}  } \right) ^{0.5}.
\end{equation}

\subsubsection{E-commerce Demand}
\begin{figure}[!ht]
\centering
\hspace{-1 cm}
\includegraphics[scale=0.5]{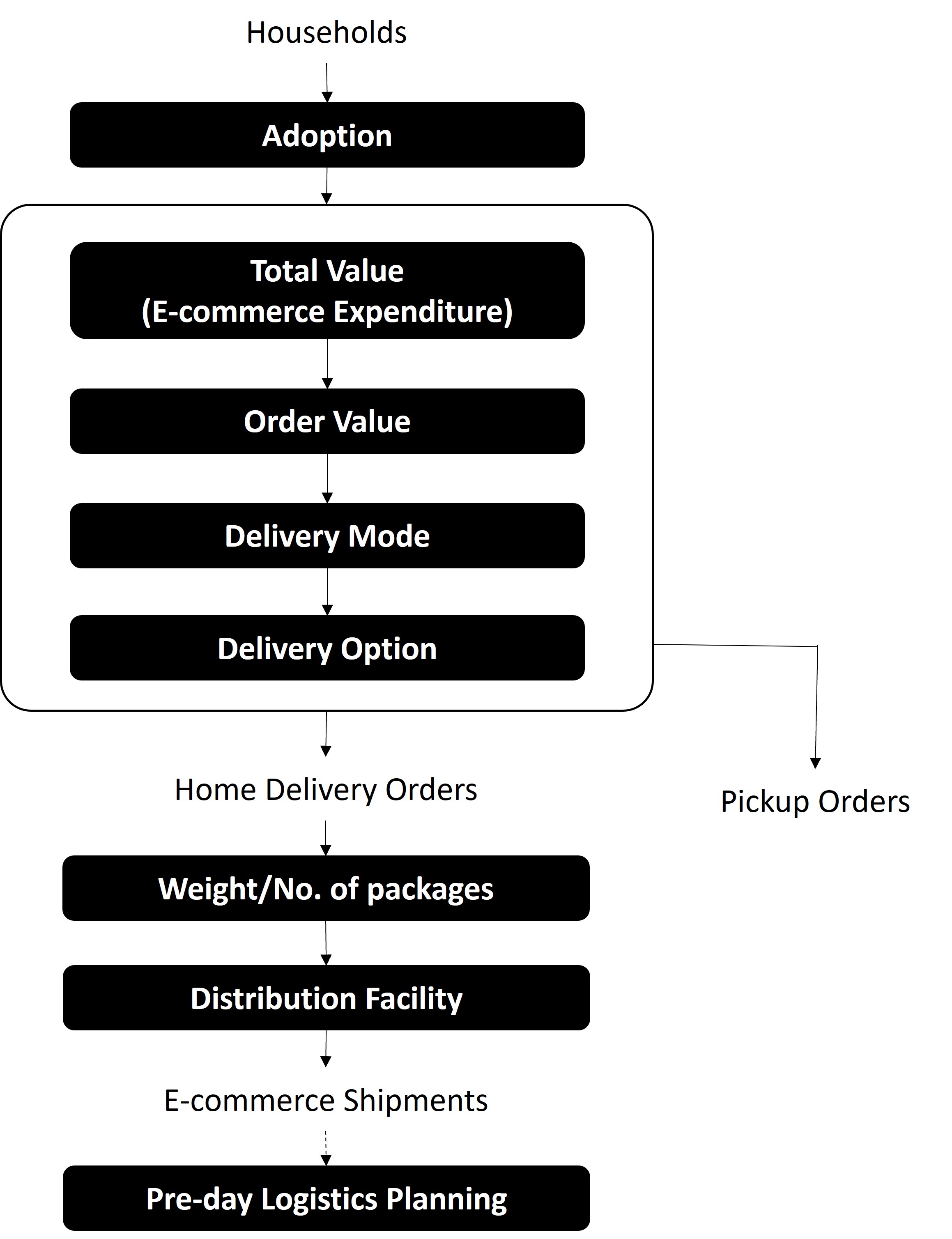}
  \caption{Flow of household-based E-commerce demand simulation. Modified from \cite{sakai2022household}}  
  \label{fig:sakai2022household}
\end{figure}

In the e-commerce demand model, each household is considered the decision-maker, which makes orders (or purchase decisions) relating to three commodity types: groceries, household goods, other durable goods. The model is a hierarchical set of discrete choice models (Figure \ref{fig:sakai2022household}). At the top level, each household decides on e-commerce adoption (on a monthly basis) based on household characteristics and the distance of off-line shopping trips by household members. If adopted, then at the second level, the household decides on the order details, which are represented by a nested logit model containing decisions on total e-commerce expenditure, order value, delivery mode, and delivery option from top to bottom. There are two delivery modes. If pickup is chosen, then the pickup option is further chosen (curbside, in-store), and a passenger pickup activity is added to the daily activity schedule of a household member. Pickup is associated with a fee. If home-delivery is chosen, the household further decides on delivery options (speed, time slot, date, etc.); each option is also associated with a fee. The order is converted to shipment(s) using simple random assignments of weight-per-value and the number of packages to each order based on calibrated, pre-defined distributions. Next, the distribution facility (i.e., origin) is assigned to each shipment using the supplier selection model. A B2C flow is thus generated, and the shipment goes to the logistics planning of shippers and carriers.

Note that fees for delivery and pickup options are designed for different categories of commodities. We assume that delivery fees to households within a tolled region will increase, although fees for the pickup option will remain the same. Although shippers might not be affected by congestion pricing or even benefit from it, we assume that they will still attempt to transfer the toll cost to receivers. We assume an increase in delivery fees by a rate such that the increase is equal to the total amount of toll paid by freight vehicles divided by the number of deliveries within the toll area (including e-commerce and non-e-commerce shipments). 

\subsubsection{Vehicle Operations Planning}
\begin{figure}[!ht]
\centering
\hspace{-1 cm}
\includegraphics[scale=0.5]{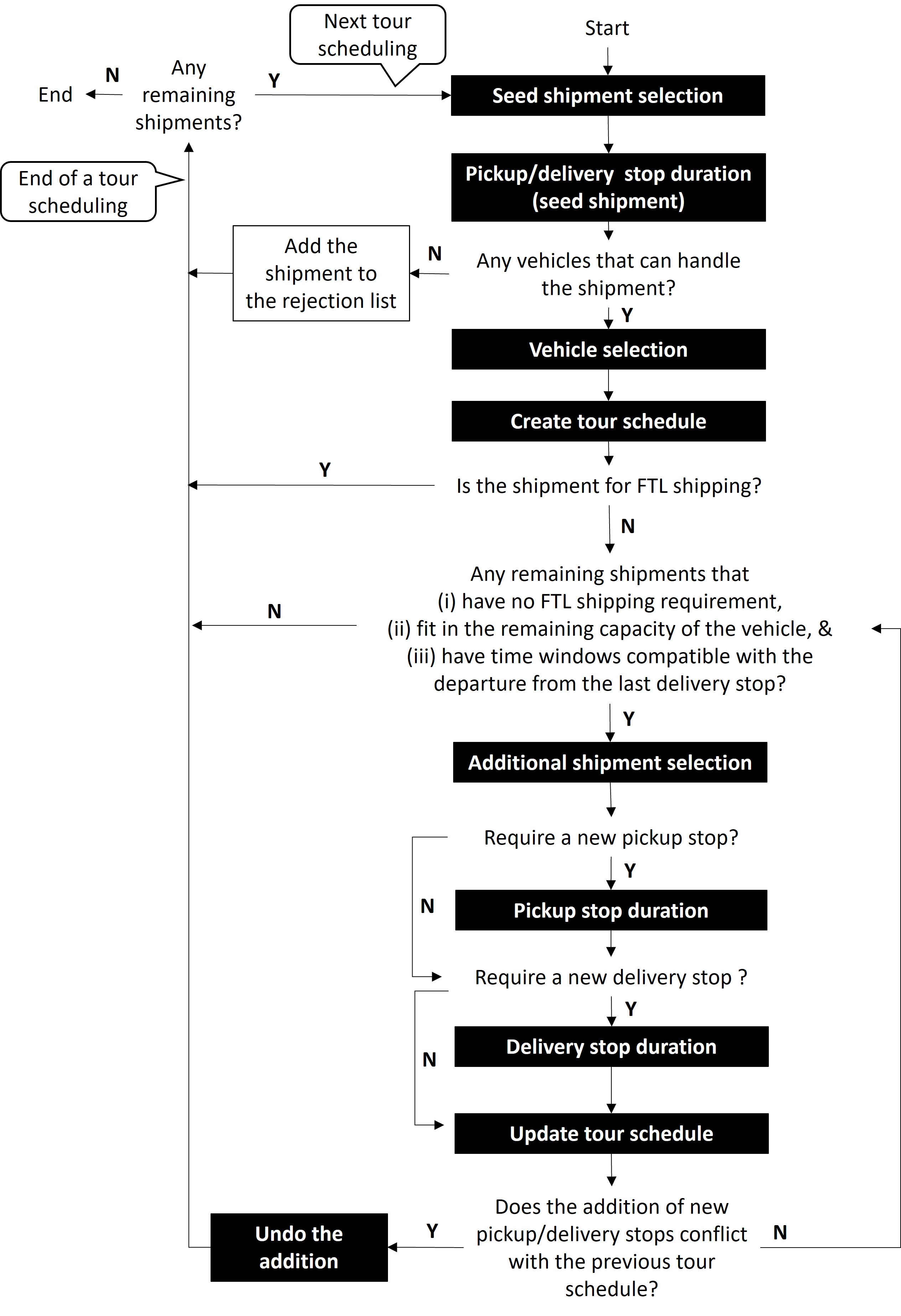}
  \caption{Flow of Vehicle Operations Planning. Modified from \cite{sakai2020simmobility}}  \label{fig:sakai2020simmobility}
\end{figure}

 The decision-makers of VOP are freight carriers. Under feasibility constraints, they make decisions on shipment-to-vehicle assignment and fleet tour schedules. The model inputs are carriers, carriers' fleets (payload capacity and body type), shipments assigned to carriers (pickup and delivery locations, delivery time window, size, and shipping requirement (parcel, full truckload, or less than truckload)), and network information. The flow of VOP proposed by \cite{sakai2020simmobility} is shown in Figure \ref{fig:sakai2020simmobility}. The heuristic assigns shipments, one-by-one, to carrier's goods vehicles and generates vehicle tours for pickups and deliveries. 
 
\begin{itemize}
  \item[1)]	Select a seed shipment ($S_0$) to assign: the one with the earliest and the narrowest pickup time window;
  \item[2)]	Draw pickup and delivery dwell time of $S_0$;
  \item[3)]	Select the vehicle with the maximum remaining working hours for $S_0$;
  \item[4)]	If $S_0$ is full truck load, terminate tour and go to 1) ;
  \item[5)]	Select the most “similar” remaining shipment ($S_1$), under the feasibility constraints;
  \item[6)]	If $S_1$’s pickup location is different from the locations already scheduled to visit, draw dwell time and add a pickup stop;
  \item[7)]	If $S_1$’s delivery location is different from the locations already scheduled to visit, draw dwell time and add a delivery stop;
  \item[8)]	If the updated tour is infeasible timewise, terminate tour and return $S_1$ to the remaining shipment list and go to 1); otherwise, repeat 5) – 8).
\end{itemize}

 The limitation of the heuristic proposed in \cite{sakai2020simmobility} is that the model is not sensitive to tolls, though it is sensitive to travel time. Therefore, we develop an MNL model to enhance the VOP model, assuming the decision rule is utility maximization under feasibility constraints. 
For a carrier $n$, the systematic part of the utility function (omitting subscript $n$) of a VOP alternative $i$ is:
\begin{equation}
\begin{aligned}\label{eqn:VOP}
&V_i= \sum_{v \in fleet}   \sum_{t \in trips_{i,v}}  -toll + \beta_1 \times time_t + \\
& (\beta_2 \times \delta_{LGV_t} + \beta_3 \times \delta_{HGV_t} + \\
& \beta_4 \times \delta_{VHGV_t} ) \times distance_t ,
\end{aligned}
\end{equation}

where $v$ is a freight vehicle, $fleet$ is the set of candidate freight vehicles belonging to the carrier, $t$ is a trip made by the vehicle, and $trips_{i,v}$ is the set of trips made by vehicle $v$ under alternative $i$. The utility function takes a money-metric form. We include toll ($toll$), total time ($time_t$), and distance ($distance_t$) as cost-related variables. Total time (including travel and dwell time at pickup and delivery stops) is used as a proxy of driver wage. We make this simple assumption because most urban freight drivers are paid by the hour, which is the most common form of payment for urban and LGV drivers. Along the lines of \cite{perera2019determining}, distance interacted with vehicle body type dummies ($\delta_{LGV_t}$, $\delta_{HGV_t}$, $\delta_{VHGV_t}$) serve as proxies for vehicle-based costs (fuel, tire, repair, maintenance, and depreciation). Carriers use travel time, distance, and toll information based on their experience from the day-to-day learning process (refer Section \ref{sec:Supply}). The travel time, distance, and toll represent values from the trip origin zone to the trip destination zone by departure time (AM peak, PM peak, or off peak).
All trips of all vehicles, including returning to overnight parking locations, are included in the cost calculation. 

Utilizing the concept of the path-size variable in route choice (\cite{ramming2001network}), we define an overlap factor to penalize the overlap between alternative plans. The idea is that the utility of an alternative similar to others in the choice set should be reduced, as suggested by the very intuitive “red bus, blue bus” paradox. To better explain it, we use an assignment matrix $A_i$ to represent each VOP alternative $i$ (omitting carrier subscript n for simplicity):
  
 \begin{equation}
A_i =
 \begin{pmatrix}
a_{1,1,i} & \cdots & a_{1,V,i}\\
\vdots & \ddots &   \vdots \\
a_{S,1,i} & \cdots  & a_{S,V,i}
\end{pmatrix}
 \end{equation}

where $a_{s,v,i}$ is an element in assignment matrix $A_i$, which takes a value 1 if shipment $s$ is assigned to vehicle $v$ and 0 otherwise, $S$ is total number of shipments to be assigned by the carrier, and $V$ is the total number of vehicles in the carrier’s fleet.

Each shipment is assigned to one vehicle, so we have,
 \begin{equation}
\sum_{v=1,...,V} a_{s,v,i} = 1, \forall s = 1,2,...S, \forall i = 1,2,...|C_n|,
 \end{equation}

where $C_n$ is the choice set of carrier $n$. 

The overlap factor (denoted $OF$) for a VOP alternative i is defined as:
  \begin{equation}
OF_i = 1 / S \times \sum_{\{s,v|a_{s,v,i}=1\}}{ 1 \over {\sum_{k=1,2,...|C_n|} a_{s,v,k}}}
 \end{equation}

The less an alternative overlaps with others in the choice set, the bigger its overlap factor is, and the maximum value it can take is 1. Conversely, the more an alternative overlaps with the others, the smaller its overlap factor, and the minimum value it can take is $1/S$.

The utility function is the sum of the systematic part, the overlap factor, and the error term, which follows an extreme value distribution with scale parameter $\mu$:
\begin{equation}\label{eqn:fuf}
U_i=V_i+\beta_5 \times \log{(OF_i)} +  \epsilon_i / \mu,
\end{equation}

In the choice set generation process, a set of feasible alternative vehicle operations plans are generated. We propose to use several alternative criteria to order the shipments and vehicles. We adapt step 1), 3), 5) to generate multiple alternative plans.

For seed shipment selection in step 1, we use four different approaches to order shipments and select the one on the top of the list as the seed.
\begin{itemize}
 \item[•]Order shipments by shipment size (decreasing order), shipping requirement (ordered as FTL, LTL, parcel), and then pickup time window (earliest delivery time first);
 \item[•]Order shipments by time window, shipment size, and then shipping requirement;
 \item[•]Order shipments by shipment size, time window, and then shipping requirement;
 \item[•]First, classify shipments into 3 groups by shipping requirement and next, order shipments by shipment size within each group.
\end{itemize}

For vehicle selection in step 3, we use two approaches to order vehicles and select the one on the top of the list: 1) order vehicles by remaining working hours (decreasing order), and 2) order vehicles by remaining capacity (decreasing order).

For measuring similarities between shipments in step 5, we use two approaches to generate the similarity matrix between shipments: 1) take the zone-to-zone distance between two pickup stops and the distance between two delivery stops, and calculate the geometric mean of these two distances, and 2) sum the difference between latest arrival times at two pickup stops and that of two delivery stops.

Each time we run the heuristic, we use one approach for steps 1, 3, and 5, respectively. Thus, we can generate a maximum of $4 \times 2\times 2=16$ alternatives for each carrier if feasibility constraints are met. 

\subsubsection{Route Choice}
The route-choice model for freight carriers is described in detail in \cite{toledo2020intercity}.

\subsection{Supply}\label{sec:Supply}
The pre-day and within-day passenger and freight models yield trip-chains for each individual and vehicle tours for each freight vehicle, which are simulated on a multimodal network using the Supply module, which is a mesoscopic traffic simulator. The Supply module includes bus and rail controllers to manage public transit operations including the frequency/headway-based dispatching of vehicles, monitoring of occupancy,
dwelling at stops and so on. More details may be found in \cite{lu2015simmobility} and \cite{oh2020assessing}.
Finally, note that demand-supply interactions are explicitly modeled through the iterative day-to-day and within-day learning processes, which involve a successive averaging of time-dependent zone-to-zone and link-level travel times respectively, until an acceptable measure of convergence between travel times in successive iterations is achieved.

\section{Prototype City Generation and Calibration}\label{sec:Sec4}

\begin{figure}[h]
\centering
\hspace{-0.5 cm}
\begin{subfigure}{.6\textwidth}
  \centering
\includegraphics[scale=0.5]{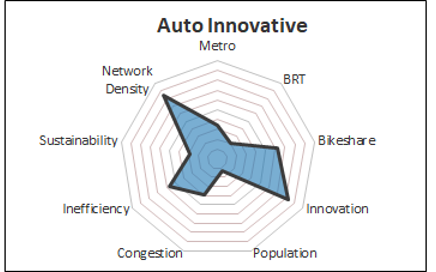}
  \caption{Auto-Innovative Characteristics}
  \label{fig:Spider_AI}
\end{subfigure}%
\hspace{0.25 cm}
\begin{subfigure}{.4\textwidth}
  \centering
  \includegraphics[scale=0.45]{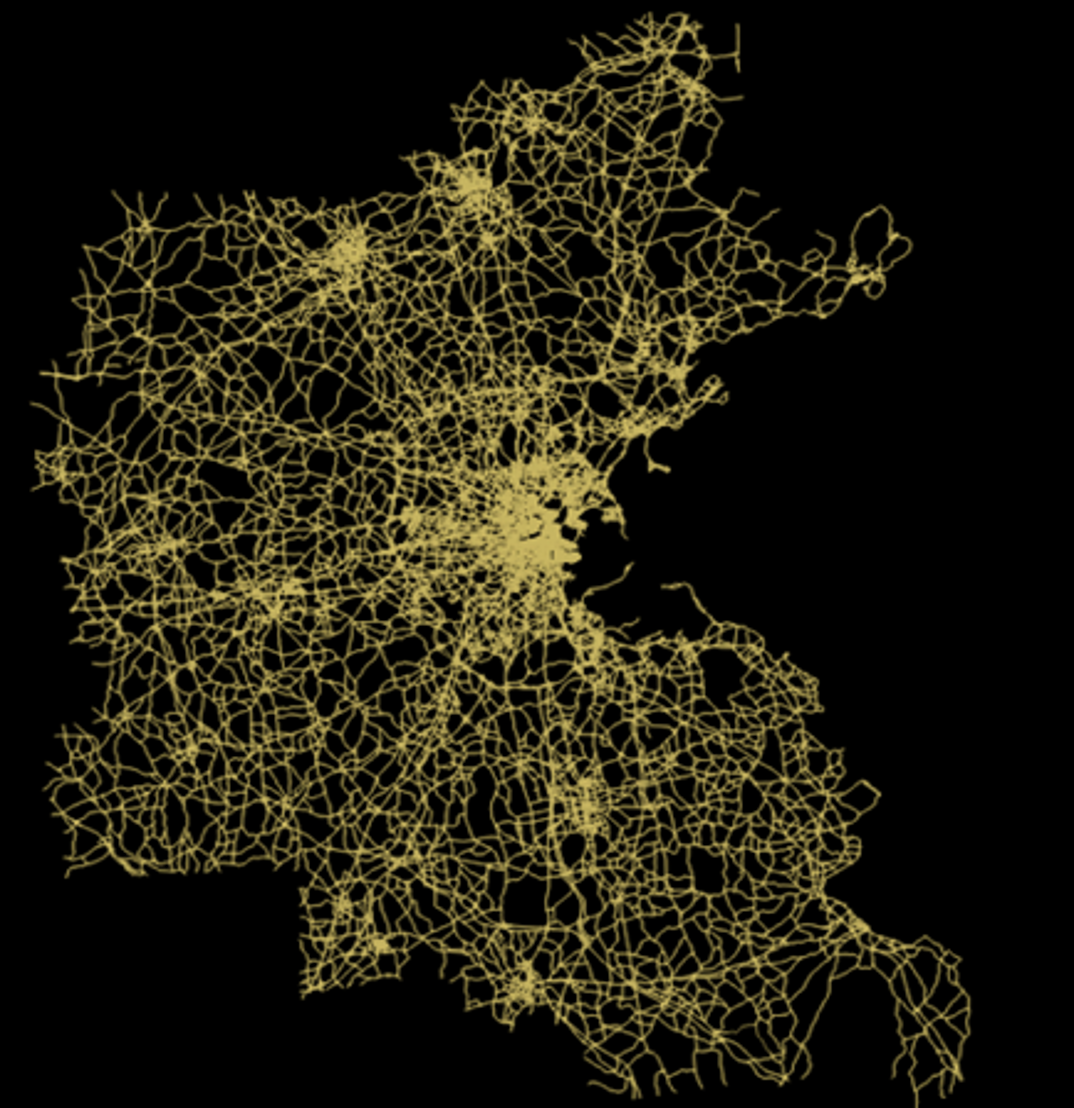}
  \caption{GBA Network}
  \label{fig:GBA_Network}
\end{subfigure}
\caption{Prototype City}\label{fig:Prototype_city}
\end{figure}

The study area is a prototype city representing a typology or class of cities termed \textit{Auto-Innovative} \citep{oke2019novel}. This type of city is characterized by modernization and industrialization, high auto-dependency, and high transit mode share, as well as high metro and population density, mainly representing North American cities (e.g., Boston, Washington D.C., Chicago, Toronto). The prototype city is synthesized based on population, land use, and supply characteristics of the Greater Boston Area (GBA), which serves as an archetype city --one that is close to typology averages on many indicators (Figure \ref{fig:Prototype_city}). The key characteristics of the city are as follows:
\begin{itemize}
\item Zoning: 164 municipalities, 2,727 traffic analysis zones (TAZs)
\item Area: 7.32 thousand square kilometers
\item Population: 1.74 million households, 4.60 million residents
\item Business establishments: 0.130 million
\item Vehicles: 2.47 million passenger vehicles, 0.378 million goods vehicles
\end{itemize}

The pre-day activity-based models were originally estimated from the 2010 Massachusetts Travel Survey data and time and cost skim matrices from Boston’s Central Transportation Planning Staff. Model specification details are described in \cite{viegas2018modeling}. These models were then calibrated to ensure that demand and activity patterns match typology averages as described in \cite{oke2020evaluating}.

\begin{table}[!h]
\caption{Aggregate cost elasticity in passenger models}\label{table:Elasticity}
\renewcommand{\arraystretch}{0.8}
\hspace{-1.5 cm}
\small
\begin{tabular}{l|l|l|l|l}
\hline
Model                                      & Level                              & Alternative                                              & Activity type         & Elasticity \\
\hline
\multirow{4}{*}{Mode-destination choice} & \multirow{4}{*}{Tour}              & \multirow{4}{*}{Car}                                     & Work                  & -0.082     \\
                                           &                                    &                                                          & Education             & -0.20      \\
                                           &                                    &                                                          & Shop                  & -0.21      \\
                                           &                                    &                                                          & Other                 & -0.23      \\
                                           \hline
\multirow{2}{*}{Mode choice}               & \multirow{2}{*}{Tour}              & \multirow{2}{*}{Car}                                     & Work                  & -0.10      \\

                                           &                                    &                                                          & Education             & -0.18      \\
                                           \hline
Mode choice                                & Sub-tour                           & Car                                                      & Work                  & -0.13      \\
\hline
Mode choice                                & Intermediate stop                  & Car                                                      & Any                   & -0.18      \\
\hline
\multirow{3}{*}{Time of day}               & \multirow{3}{*}{Tour}              & \multirow{3}{*}{Depart during AM or   PM peak using car} & Work                  & -0.11      \\
                                           &                                    &                                                          & Education             & -0.15      \\
                                           &                                    &                                                          & Shop/Other            & -0.23      \\
                                           \hline
\multirow{3}{*}{Time of day}               & \multirow{3}{*}{Sub-tour}          & Depart during AM   peak using car                        & \multirow{3}{*}{Work} & -0.14      \\
                                           &                                    & Depart during PM   peak using car                        &                       & -0.16      \\
                                           &                                    & Depart during off   peak using car                       &                       & -0.21      \\
                                           \hline
\multirow{3}{*}{Time of day}               & \multirow{3}{*}{Intermediate stop} & Depart during AM peak   using car                        & \multirow{3}{*}{Any}  & -0.18      \\
                                           &                                    & Depart during PM   peak using car                        &                       & -0.19      \\
                                           &                                    & Depart during off   peak using car                       &                       & -0.23     \\
                                           \hline
\end{tabular}

\end{table}

Further, to better represent heterogeneity in travelers, we introduced lognormally distributed (see \cite{hess2005estimation}) values of time (VOT) to the time-of-day and mode-destination choice models, where individuals from higher income groups are assumed to have a higher VOT. The mean-values of the parameters associated with the cost variables were calibrated to ensure that the aggregate cost elasticities (Table \ref{table:Elasticity}) accord with empirical values indicated in the literature (mode choice: \cite{vega2008employment,bhat2000incorporating,yang2013cross}; time-of-day choice: \cite{ding2015cross,sasic2013modelling,bhat1998analysis}). As can be seen, trips and tours associated with work activities are less elastic than shopping and other discretionary activities. Note that in order to determine the standard deviation of the randomly distributed cost co-efficients, we conservatively assumed a co-efficient of variation of 0.2 (the literature reports a ranges of values, see \cite{seshadri2022congestion}). 

The prototype city generation described in \cite{oke2020evaluating} considers the passenger side only. In order to model freight, we augmented this synthetic prototype city by synthesizing business establishments. An establishment has attributes including a location, industry type, function type, employment size, floor area, fleet, and drivers. The role of an establishment can be a receiver, a shipper, a carrier, or a combination of these. Each establishment is treated as an independent agent in the simulation.

 \begin{figure}[h]
\hspace{-1 cm}
\includegraphics[scale=0.75]{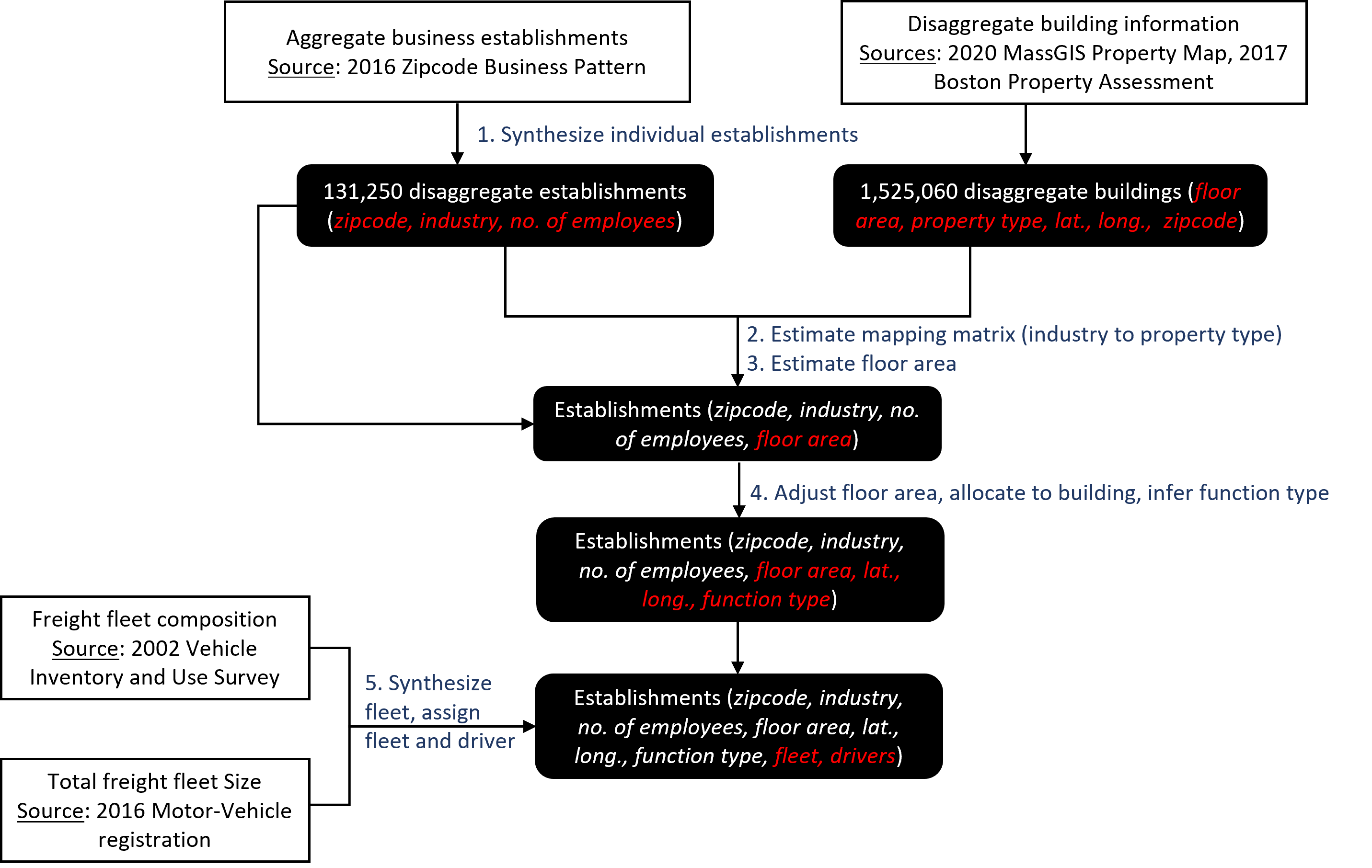}
  \caption{Synthesis of Business Establishments}  \label{fig:EstSynthesis}
\end{figure}

The proposed procedure for the synthesis of business establishments is summarized in Figure \ref{fig:EstSynthesis} and relies exclusively on publicly available data. Specifically, four different data sets are used. The first (dataset 1) is aggregate establishment information, including industry type and employment size. Although disaggregate or micro-samples of business establishments would be ideal, to the best of our knowledge, there are no open-source datasets of these. Therefore, we use the 2016 ZIP Code Business Patterns (ZBP) for the GBA published by the U.S. Census Bureau (2016). Only county-level data are published by the Census Bureau since 2017, so we use the 2016 dataset as it is more disaggregate. ZBP contains the total number of business establishments classified by the North American Industry Classification System (NAICS) code and by employee size category at the zip code level, corresponding to industry type and employment attributes. The second dataset (dataset 2) contains disaggregate geographical information of buildings. Compared with disaggregate establishment data, disaggregate building information is much easier to obtain. Example sources include OpenStreetMap and property tax reports, publicly available for most North American cities. We use 2020 MassGIS Massachusetts Interactive Property Shapefiles \footnote{https://www.mass.gov/service-details/massachusetts-interactive-property-map} which covers the GBA except for Boston, and the 2017 Boston Property Assessment report \footnote{https://data.boston.gov/dataset/property-assessment} for Boston. This dataset contains records of each building, including its latitude, longitude, municipality, property type code, floor area, etc. The third dataset (dataset 3) is the total number of freight vehicles, usually available from published vehicle registration reports. For the GBA, we obtain total freight fleet size from the 2016 State Motor-Vehicle Registrations of Massachusetts \footnote{https://www.fhwa.dot.gov/policyinformation/statistics/2016/mv1.cfm}. Finally, the fourth dataset (dataset 4) is the proportions of freight vehicles by classes such as industry and body type. We use the 2002 Vehicle Inventory and Use Survey report for Massachusetts (U.S. Census Bureau, 2002). This survey has been resumed in recent years, so more updated information may become available soon.

The five steps in the business establishment synthesis (Figure \ref{fig:EstSynthesis}) are as follows:
\begin{enumerate}
    \item Generate individual establishments from dataset 1 by randomly and uniformly sampling employment size within the corresponding class. Each establishment has a zip code, industry, and employment size attributes.
    \item 	Estimate a mapping matrix $\boldsymbol{X}$ between industry code and property code, where the row is industry code, and the column is property code. Entry $x_{ij}$ is a decision variable to be solved for, where $i$ is an industry code, $j$ is a property code. We map the 24 two-digit NAICS codes to the 11 industry types in SimMobility: chemical manufacturing, metal manufacturing, machinery manufacturing, light manufacturing, road freight transportation service, water/air freight transportation service and transportation-related service, warehousing, material wholesale, product wholesale, retail, and restaurant and service. We identify 25 unique property codes that can accommodate establishments and select the buildings feasible for assignment. The city has 282 zip code areas with at least one feasible building and one establishment that is not a P.O. box. The problem of determining $\boldsymbol{X}$ is formulated as an integer programming problem (Appendix A) with the objective to minimize the number of feasible mappings, i.e., to make the matrix as sparse as possible to reduce complexity in the establishment-to-building assignment. 
    \item Estimate the floor area of each establishment assuming it is proportional to employment size up to a conversion factor plus a constant (as in \cite{le2016constructing}). The conversion factor $r_{ik}$ and constant $p_{ik}$ are both specific to industry $i$ and zone (zip code) $k$. We formulate a non-linear programming problem (Appendix B) to determine $r_{ik}$ and $p_{ik}$ with the objective to minimize the difference between total establishment floor area and total building floor area for each zip code and each property type (to maximize the utilization of each building’s floor area). Following this, the estimated floor area of an establishment $m$ of industry type $i$ in zip code $k$, denoted by  $\hat{f}_{ikm}$ is given by $\hat{f}_{ikm} = r_{ik}N^E_{ikm} + p_{ik} $, where $N^E_{ikm}$ is the number of employees in establishment $m$ of industry type $i$ in zip code $k$. Thus, each establishment now has the attributes zip code, industry, employment size, and floor area. 
    
    \item Adjust establishment floor area, assign each establishment to a building, and determine its function type. The floor area adjustment is needed to ensure the buildings in each zip code $k$ have enough floor area for accommodating the assigned establishments. The readjustment of the floor areas is once again formulated as a non-linear programming problem (similar to that in Appendix B) for each zip code $k$. We omit the details for the sake of brevity. Following this, adjusted floor areas are calculated. 
    Next, we assign establishments to buildings. We formulate an integer programming problem to minimize the difference between the building floor area and the total floor area of the assigned establishments (Appendix C) .
    Lastly, we generate function type for establishments based on the industry type and building property type combination using judgement and marginal probabilities from Singapore \citep{sakai2020simmobility}. In SimMobility, the five establishment function types are office, factory, retail and restaurant, logistics facility, and others. Note that for some industries, certain function types do not exist, e.g., a warehousing establishment cannot have a function of factory or  retail and restaurant. Thus, the establishments now have attributes of zip code, industry type, function type, employment size, floor area, latitude, and longitude.
    
    \item Generate freight vehicle fleets and assign them to establishments. From dataset 4, we obtain the total number of freight vehicles by maximum laden weight (MLW) (light goods vehicle – LGV, heavy goods vehicle – HGV, very heavy goods vehicle – VHGV) and the number by industry type for the year of 2002. We apply a growth rate to the 2002 statistics to match the total size of the freight fleet in 2016, assuming equal growth rates for different vehicles and industries. We first perform iterative proportional fitting (IPF) to generate the total number of each type of vehicles for each industry type. The target margins of the IPF matrix are the total number of vehicles by type (3 rows) and the total number by industry owned (11 columns) in 2016. After this, we generate a vehicle fleet, each with attributes of type and industry owned.
    
    Next, we assign vehicles to each establishment based on its industry type, function type, and employment size. We first generate a matrix $\boldsymbol{Z}$ for each vehicle type where the rows are function types (5) and columns are industry types (11), representing a prior belief for the split of the vehicles between different establishment function types. An element of the matrix $z_{f,i,v}$ is a ratio of the number of vehicles of type $v$ owned by establishments of industry $i$ and function  $f$ to that owned by establishments of industry type $i$:
        \begin{equation}
        z_{f,i,v} = \frac{\sum_m N^V_{m(i,f)}}{\sum_m N^V_{m(i)} },
        \end{equation}
    Given that detailed data on fleet compositions is unlikely to be publicly available, one needs to use judgement to generate the ratios or depend only on industry type and employment size. We use statistics from Singapore \citep{sakai2020simmobility} as a reference in generating these ratios. We can now compute the total fleet size for each industry and function type as $N^V_{f,i,v} = z_{f,i,v} N^V_{i,v} $.
    
    Finally, vehicles are assigned by type to establishments of each industry and function combination, assuming that the number of vehicles varies linearly with employment size. The number of vehicles of type $v$ for an establishment of industry $i$ and function type $f$ is assumed to be:
    \begin{equation}
    N^V_{m(i,f),v} = \left( \frac{N^E_{m(i,f)}}{\sum_i N^E_{m(i,f)} }\right) N^V_{f,i,v}
    \end{equation}
    
    Without sufficient data about drivers, we simply assign one driver to each vehicle. Although in practice multiple drivers may operate one vehicle and one vehicle may be shared among multiple drivers, our assumption does not essentially affect models designed in SimMobility. The driver type is for-hire if the establishment has more than one employee and is an owner operator otherwise. Thus, the establishments now have all attributes required for the SimMobility freight models: zip code, industry type, function type, employment size, floor area, latitude, longitude, fleet, and drivers. 
\end{enumerate}

We also calibrated the long-term model parameters, which determine B2B commodity flows. The initial long-term model parameters were estimated using data from the 2013 Tokyo Metropolitan Freight Survey. We re-calibrated them based on a Public Use Microdata (PUM) from the 2012 Commodity Flow Survey (CFS) (U.S. Census Bureau, 2012), which contains shipment records including origin, destination, size, commodity type, and shipper industry type. The samples in PUM are mainly manufactures and wholesalers. We also used the aggregate commodity OD flows from Freight Analysis Framework (FAF4) data to complement to PUM for other industry types. We introduced adjustment factors for production model parameters so that, for manufacturers and wholesalers, the aggregated commodity production by industry and commodity type from the model correspond to those based on PUM. After this, we introduced adjustment factors for production and consumption model parameters (except those already adjusted) to match the aggregated simulated production and consumption (by commodity type) with those based on FAF4. Further, the shipment size model parameters (Equation \ref{eqn:SMS}) are also calibrated to match the cumulative distributions of simulated shipment size for manufacturing and wholesale industries with those from on PUM. 

To calibrate e-commerce demand model, various types of datasets were used while it counted mainly on 2017 National Household Travel Survey (U.S. Federal Highway Administration) data, which include the count of times purchased online for delivery. The details of the calibration process are available in \cite{sakai2022household}.

For setting the model parameters for the VOP model ($\beta$s and $\mu$ shown in Equation \ref{eqn:VOP} and \ref{eqn:fuf}), we used various sources. The parameter values and sources are summarized in Table \ref{table:param_vop}. As the prototype city is representative of North American cities, we mainly refer to documents published by the American Trucking Research Institute (ATRI), which give factors for converting time and distance to itemized costs. However, over 80\% of the vehicles studied by ATRI are intercity trucks operating at higher speeds than urban vehicles. Therefore, we used the speed to fuel consumption relationship of freight vehicles operating at lower speeds in urban areas \citep{nilim2016}; we adjusted the distance-to-fuel-cost factors in the ATRI report \citep{ATRI2020} for the urban setting. Next, the coefficient of the overlap factor and the scale parameter are affected by the level of stochasticity in decision-making and are not easily found in literature. We used the truck driver survey data in North America conducted in 2012 \citep{ben2016freight} to obtain the estimates for them as well as the route choice model parameters. The route choice model is an MNL model. The dataset and choice set generation process for the route choice model are detailed in \cite{toledo2020intercity}, and the specification of the utility function used in SimMobility is available in \cite{sakai2020simmobility}. For estimating the overlap factor and the scale parameter ($\beta_5$ and $\mu$), we used the specification of the route choice model but replaced the route’s path size with the VOP overlap factor.

\begin{table}[]
\caption{Parameter values in the VOP model}\label{table:param_vop}
\begin{tabular}{p{0.1\linewidth}p{0.1\linewidth}p{0.8\linewidth}}
\hline
Parameter & Value & Sources \\ \hline
$\beta_1$ & -27.32 & Hourly driver wage and benefit in Murray and Glidewell (2020); cross-validated with Massachusetts average delivery driver wage \\ \hline
$\beta_2$ & -0.57 & \multirow{3}{10cm}{Based on \cite{ATRI2020}, fuel consumption rates adjusted to urban speed using \cite{nilim2016}} \\
$\beta_3$ & -0.67 &  \\
$\beta_4$ & -0.72 &  \\ \hline
$\beta_5$ & 1.61 & \multirow{2}{10cm}{Estimated values from truck route choice model with data from  \cite{ben2016freight} with the same specification} \\
$\mu$ & 1.18 &  \\ \cline{1-3}
\end{tabular}
\end{table}

\section{Design of Congestion Pricing Schemes}\label{sec:Sec5}
In this section, we discuss the overall scenario design, the definition of the tolling zone or area, the definition of toll rates and the performance measures that will be used to evaluate the pricing schemes.

\subsection{Scenario Design}\label{sec:Sec5_1}
We consider the following five scenarios:
\begin{itemize}
    \item Baseline scenario or do-nothing (Base)
    \item Congestion Pricing
        \begin{itemize}
        \item Distance-based pricing (Distance): toll charge is proportional to the distance traveled within the tolled area with an upper limit for each vehicle during a day. The tariff is time-varying. 
        \item Cordon-based pricing (Cordon): a vehicle is charged each time it enters the tolled area through radial road links. Vehicles exiting the area are not tolled. The tariff is time-varying.
        \item Area-based pricing (Area): a vehicle traveling within the tolled area from the start of the AM peak to the end of the PM peak pays a flat toll charge during a day.
        \end{itemize}
\end{itemize}


The congestion pricing policies we examine share several features. First, they are \textit{location-specific} in that tolls are applied for a specific area of the city consisting of several traffic analysis zones. Second, they are \textit{time-period-specific} in that toll-rates are different during different periods of the day. Third, they are \textit{vehicle-type} specific in that toll-rates are based on passenger-car-units of the vehicle. Based on the maximum laden weight, goods vehicles are classified as light goods vehicles (LGV) (no more than 3.5 ton), heavy goods vehicles (HGV) (3.5-16 ton), very heavy goods vehicles (VHGV) (more than 16 ton), of which the PCU is 1.5, 2, 2.5 respectively.

 \begin{figure}[h]
\centering
\includegraphics[scale=0.75]{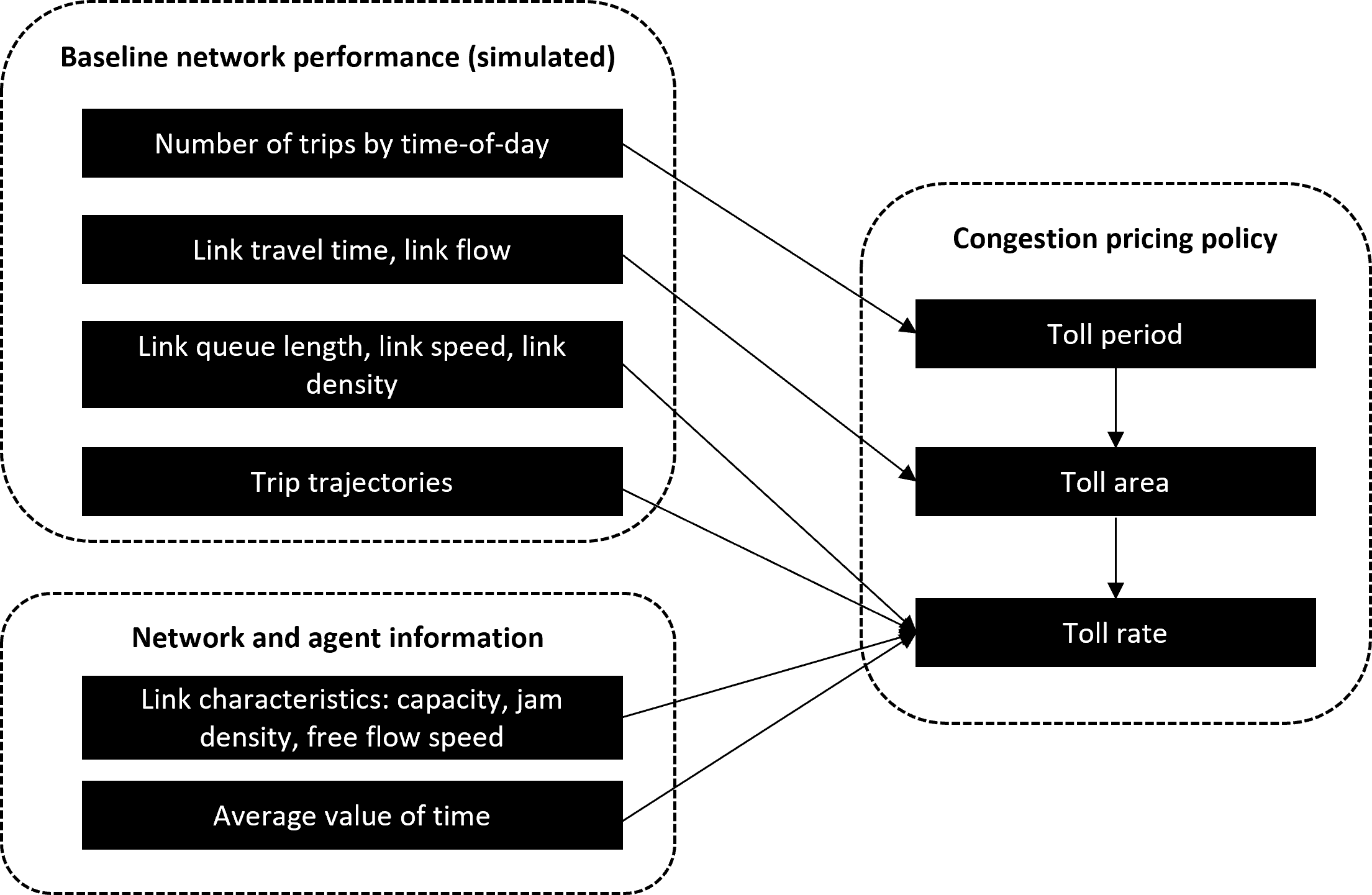}
  \caption{Framework for congestion pricing policy design} \label{fig:Pricing_design}
\end{figure}

The overall framework adopted for the design of the pricing schemes is summarized in Figure \ref{fig:Pricing_design}. The design utilizes information of network performance from a simulation of the baseline scenario along with agent-level information. We first examine the number of trips by time-of-day to identify the cities' peak hours and determine toll period(s) accordingly. Next, we quantify the extent of congestion during the tolling periods at the level of traffic analysis zones (TAZ) to define a tolling area for all three pricing schemes. Finally, we use simulated network performance information to approximately estimate congestion externalities of each link (and for all tolled vehicles using trajectories) by time-of-day. The externalities are then monetized using an average value of time to determine the toll rates. Details of each step are provided in the following sections. 

 \begin{figure}[h]
\hspace{-1 cm}
\begin{subfigure}{.5\textwidth}
  \centering
\includegraphics[scale=0.85]{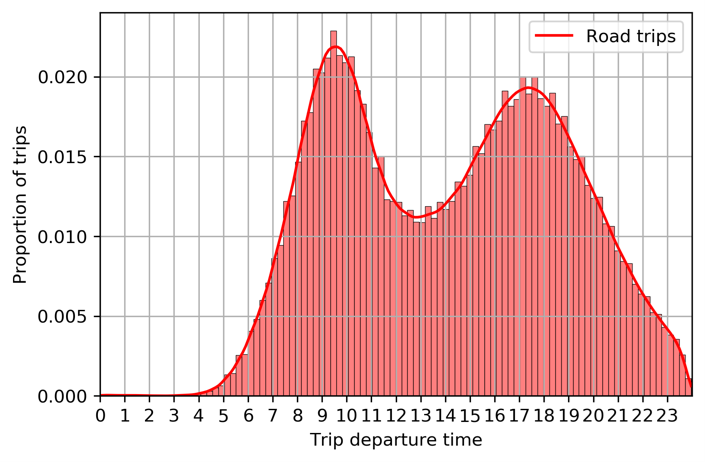}
  \caption{Passenger car trips by time-of-day}
  \label{fig:Pass_TOD}
\end{subfigure}%
\hspace{1.0 cm}
\begin{subfigure}{.5\textwidth}
  \centering
  \includegraphics[scale=0.85]{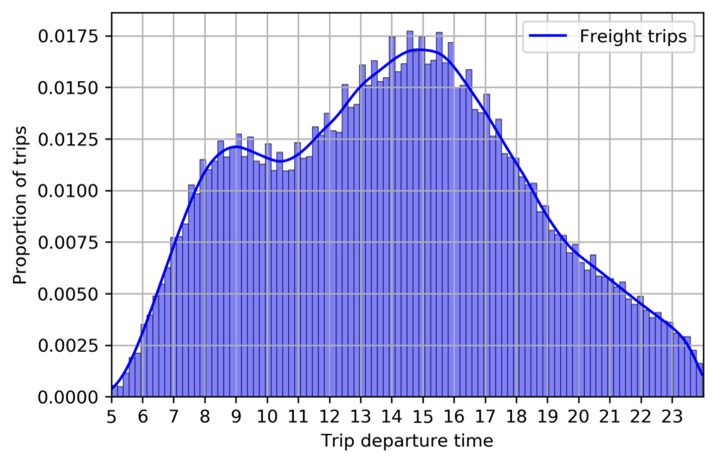}
  \caption{Freight trips by time-of-day}
  \label{fig:Freight_TOD}
\end{subfigure}  
   \caption{Trip time-of-day patterns} \label{fig:Trips_TOD}
\end{figure}

\subsection{Toll Period}\label{sec:Sec5_3}
Figure \ref{fig:Trips_TOD} shows the number of passenger car and freight trips by departure time on an average weekday from the calibrated model. Based on these, we define tolling periods (for the distance- and cordon-based schemes) between 8-10 AM (AM peak) and 4-7 PM (PM peak). Tolls are also applied during 30 minutes before and after the peak periods with lower rates, as ramp up/down periods to avoid abrupt changes in demand. It is worth-noting that the freight trips have significantly different patterns of departure time. However, given that the overall proportion of freight trips is small ($\sim 7\%$), we define the tolling periods largely based on passenger trip patterns. 

\subsection{Toll Area}\label{sec:Sec5_2}
In order to define the toll area, we use the travel time index (TTI), which is the ratio congested travel time to free-flow travel time as an indicator of congestion. The link-flow weighted average TTI is computed across all links for each TAZ. Based on the computed TTI, we can identify an appropriate area or region to be tolled. There are several additional considerations based on judgment. The tolled area should cover as many congested TAZs as possible and natural barriers (e.g., rivers, coasts, freeways) may help define borders of the tolled region for ease of implementation.

\begin{figure}[h]
\begin{subfigure}{.5\textwidth}
  \centering
\includegraphics[scale=0.5]{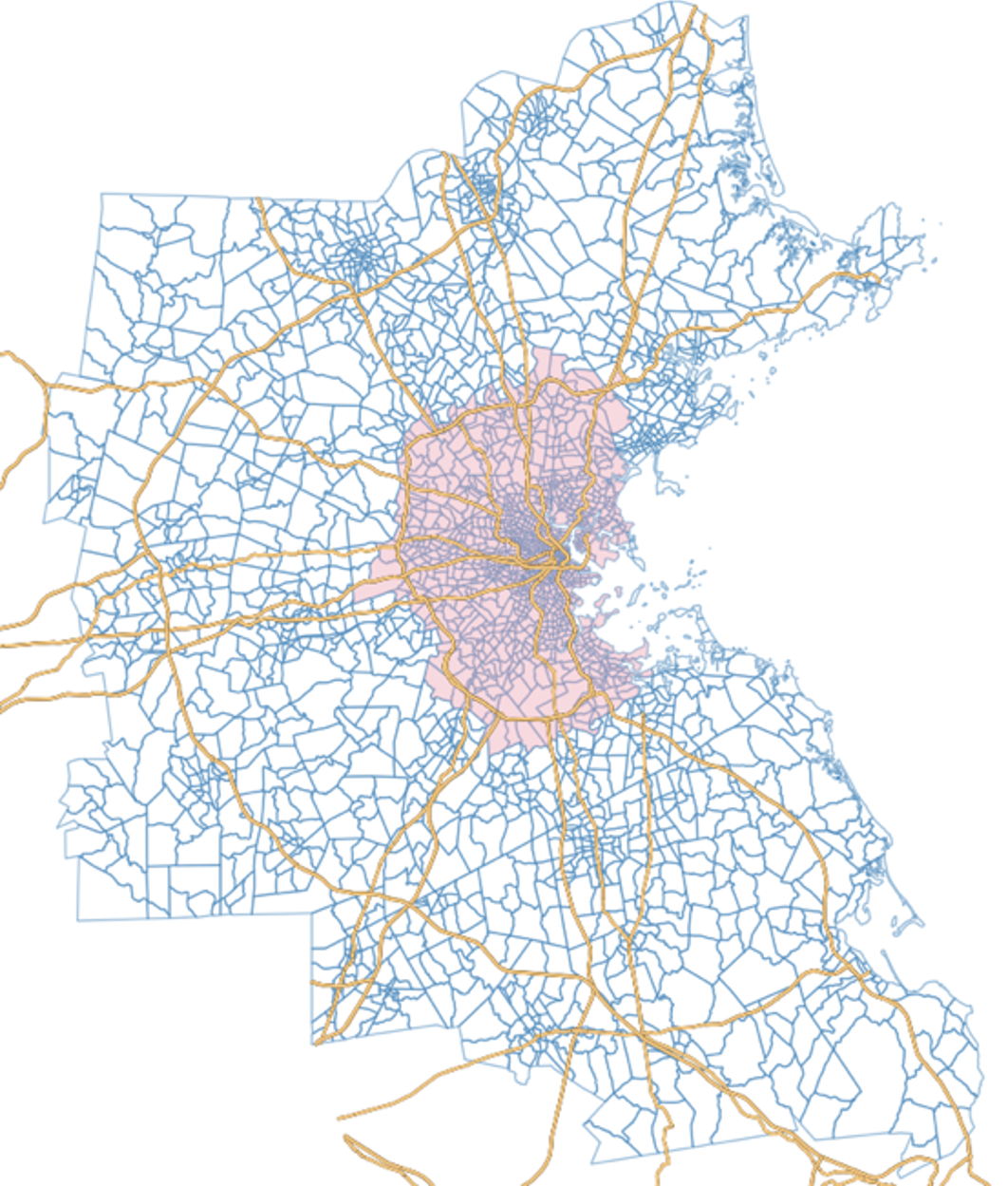}
  \caption{Tolled area}
  \label{fig:Tolled_area}
\end{subfigure}%
\begin{subfigure}{.5\textwidth}
  \centering
  \includegraphics[scale=0.5]{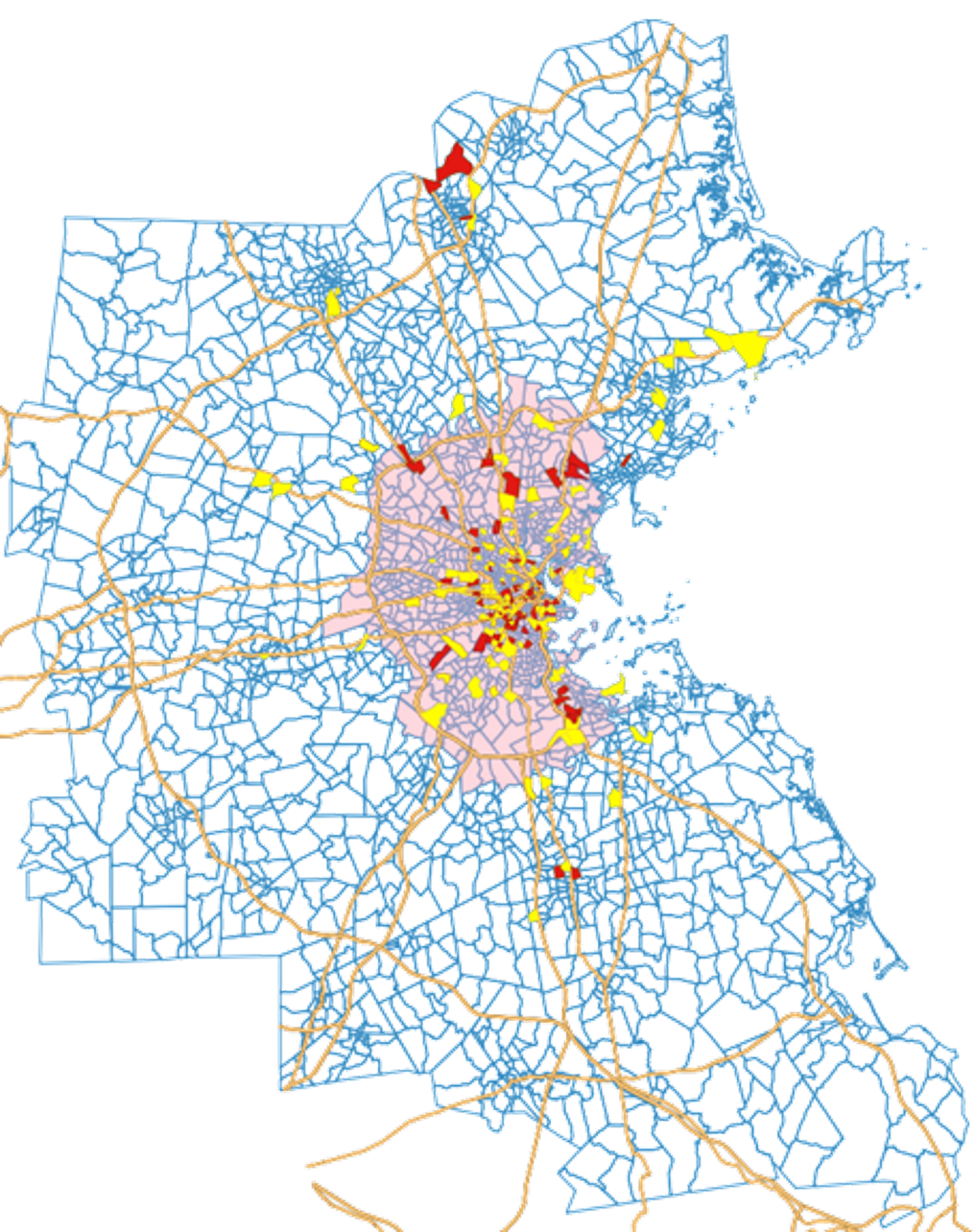}
  \caption{TAZs with high TTI}
  \label{fig:TAZ_TTI}
\end{subfigure}  
   \caption{Tolled area in prototype city} \label{fig:TollA}
\end{figure}

As shown in Figure \ref{fig:TAZ_TTI}, we find the TAZs with a relatively higher average TTI (1.2+) are highly concentrated in the eastern bay of the city (yellow indicates zones with a high TTI in the AM peak while red zones indicate a high TTI in the PM peak). We define the tolled area as the shaded area so it includes most of the those TAZs. The eastern boundary is the natural coastline and the western boundary of is a major freeway. The toll area spans over 791 square kilometers (10.8\% of total area) and has 1.85 million residents (40.2\% of total population) and 9,652 business establishments (7.7\% of total), which indicates that the population density is much higher than that of the entire city.

The temporal and spatial patterns of trips in the Base scenario are summarized in Table \ref{table:total_trips}, based on trip departure time and origin-destination (OD) pair. External trips have both ODs outside of the tolled area, internal trips have both ODs within the tolled area, and connection trips have the destination within the tolled area (entering) or origin within the tolled area (exiting). Among passenger trips including all modes, 32.9\% are internal to the toll area, 49.3\% are external, and 17.8\% are connection. Among freight trips, 34.1\% are internal, 47.1\% are external, and 18.8\% are connection. Within external trips, a very small proportion pass through the tolled area, counting for about 1.2\% of all passenger trips and 0.5\% of all freight trips. 

\begin{table}[]
\caption{Total trips: Passenger and freight}\label{table:total_trips}
\renewcommand{\arraystretch}{0.85}
\hspace{-0 cm}
\begin{tabular}{lllllll}
\hline
Time-of-day              & Trip type                     & Total & Internal & Entering & Exiting & External \\
\hline
\multirow{4}{*}{All Day} & Passenger   trips (million)  & 12.1  & 3.99     & 1.08     & 1.07    & 5.97     \\
                         & \% in   Passenger trips       & 100\% & 32.9\%   & 8.9\%    & 8.8\%   & 49.3\%   \\
                         & Freight   trips (million)     & 1.00  & 0.34     & 0.09     & 0.09    & 0.47     \\
                         & \% in   Freight trips         & 100\% & 34.1\%   & 9.4\%    & 9.3\%   & 47.1\%   \\
                         \hline
\multirow{4}{*}{AM Peak} & Passenger   trips (million)  & 2.01  & 0.66     & 0.22     & 0.15    & 0.98     \\
                         & \% in   Passenger trips       & 100\% & 32.8\%   & 11.1\%   & 7.3\%   & 48.8\%   \\
                         & Freight   trips (million)     & 0.11  & 0.03     & 0.01     & 0.01    & 0.06     \\
                         & \% in   Freight trips         & 100\% & 27.3\%   & 8.8\%    & 9.3\%   & 54.5\%   \\
                         \hline
\multirow{4}{*}{PM Peak} & Passengers   trips (million) & 2.84  & 0.93     & 0.25     & 0.26    & 1.40     \\
                         & \% in   Passenger trips       & 100\% & 32.7\%   & 8.8\%    & 9.2\%   & 49.3\%   \\
                         & Freight   trips (million)     & 0.44  & 0.15     & 0.05     & 0.04    & 0.20     \\
                         & \% in   Freight trips         & 100\% & 34.1\%   & 11.2\%   & 9.3\%   & 45.5\% \\
                         \hline
\end{tabular}
\end{table}

\subsection{Toll Rates}\label{sec:Sec5_3}
In order to set the toll rates for each charging scheme, we approximately estimate the marginal external costs of congestion (i.e., the marginal cost tolls) using network performance measures from the baseline simulation. Thus, they can be viewed as \textit{third-best} pricing schemes (i.e., we use first-best prices in a second-best setting, see \cite{de2005comparison}). Within the mesoscopic supply model, each network segment consists of a moving part (where traffic dynamics are governed by a modified Greenshields model) and a queuing part (where vehicles incur deterministic queuing delays). For a given five-minute time interval, the marginal cost toll on a segment $i$, $\delta_i$ can be expressed as (see \cite{lentzakis2020hierarchical} and \cite{yang1998principle}):
\begin{equation}\label{eqn:MCT}
\delta_i = \gamma q_i \left(\frac{dt_i}{dq_i}\right) + \gamma \left(\frac{n^q_i}{c_i}\right),
\end{equation}
where $\gamma$ is an average value of time of travelers traversing the segment in the five-minute interval in \$ per hour, $t_i$ is the average travel time on segment $i$, $n^q_i$ is the average number of queuing vehicles, $q_i$ is the flow in vehicles per hour during the five-minute interval and $c_i$ is the segment capacity in vehicles per hour. The expression for the first term may be found in \cite{lentzakis2020hierarchical} and the second term represents the average queuing delay of a vehicle during the five-minute interval. Link-level MCTs can be computed by summing up the associated segment-level MCTs.

From Equation \ref{eqn:MCT}, we can compute the total MCT for a given trip within the tolled area, by summing up the link-level MCTs (computed at the time-interval corresponding to entry on the link) for all component links on the path within the tolled region. In the case of the distance-based tolling policy, we can compute the toll tariff per unit distance as the total MCT for all tolled trips in the specific time period (for example, AM peak) divided by the total trip distance in this period. Similarly, in case of the cordon-based policy the toll rate is computed as the total MCT of tolled trips divided by the number of tolled trips. Finally, for the area-based scheme, the toll rate is the total MCT divided by the total number of vehicles traveling in the specific time-period within the region. 
 
\begin{table}[]
\caption{Toll rates for distance- and cordon-based schemes}\label{table:toll_rates}
\renewcommand{\arraystretch}{0.85}
\hspace{-0 cm}
\begin{tabular}{l|llll|llll}
\hline
\multirow{2}{*}{Entrance time} & \multicolumn{4}{l|}{Distance-based toll (USD/km)} & \multicolumn{4}{l}{Cordon-based toll (USD/entry)}            \\
                               & Car        & LGV        & HGV        & VHGV      & Car  & LGV       & HGV       & VHGV      \\
                               \hline
07:30-07:55                    & 0.16       & 0.24       & 0.32       & 0.4       & 1.65 & 2.50      & 3.25      & 4.05      \\
07:55-08:00                    & 0.26       & 0.38       & 0.51       & 0.64      & 2.60 & 3.90      & 5.20      & 6.50      \\
08:00-10:00                    & 0.32       & 0.48       & 0.64       & 0.8       & 3.25 & 4.90      & 6.50      & 8.10      \\
10:00-10:05                    & 0.26       & 0.38       & 0.51       & 0.64      & 2.60 & 3.90      & 5.20      & 6.50      \\
10:05-10:30                    & 0.16       & 0.24       & 0.32       & 0.4       & 1.65 & 2.50      & 3.25      & 4.05      \\
15:30-15:55                    & 0.14       & 0.2        & 0.27       & 0.34      & 1.50 & 2.25      & 3.00      & 3.25      \\
15:55-16:00                    & 0.22       & 0.32       & 0.43       & 0.54      & 2.50 & 3.60      & 4.80      & 6.00      \\
16:00-19:00                    & 0.27       & 0.41       & 0.54       & 0.68      & 3.00 & 4.50      & 6.00      & 7.50      \\
19:00-10:05                    & 0.22       & 0.32       & 0.43       & 0.54      & 2.50 & 3.60      & 4.80      & 6.00      \\
19:05-19:30                    & 0.14       & 0.2        & 0.27       & 0.34      & 1.50 & 2.25      & 3.00      & 3.25   \\
\hline
\end{tabular}
\end{table}

The design toll rates based on the computed MCTs from the baseline simulation are summarized in Table \ref{table:toll_rates} for the distance- and cordon-based schemes. We employ a step-toll profile similar to that implemented in Stockholm and Singapore. The two shoulders before and after the peak-period avoid an abrupt discontinuous increase in the toll from 0 to the peak-period value. We assume that during a five minute interval before and after the peak-period (for example, 07:55-08:00 and 10:00-10:05 in the morning peak), the toll rate is 0.8 times the peak-period rate. Furthermore, during the 25-30 minute interval prior to and following the peak-period (for example, 07:30-07:55 and 10:05-10:30 in the morning peak), the toll rate is 0.5 times the peak-period rate.

For the area-based scheme, the tolls are computed to be $2.65, 4, 5.5$ and $6.6$ USD for cars, LGVs, HGVs, VHGVs, respectively. The tolls are flat and apply to any vehicle traveling within the area between the period from 8 AM to 7 PM.

The design tolls rates for the cordon-based scheme appear reasonable in comparison with several examples from practice:
\begin{itemize}
    \item ERP, Singapore: 0.37-4.41 USD
    \item Milan Area C scheme, Italy: 2.35-5.87 USD
    \item Motorway Essingeleden from Solna to Stockholm, Sweden: 2.27-3.40 USD
    \item SR520 serving downtown Seattle: 3.40-4.30 USD
\end{itemize}

For distance-based pricing policies, the few examples from practice in an urban setting largely target long-haul heavy vehicles, for example:
\begin{itemize}
\item Germany, motorways, 7.5 – 11.99 tons: 0.11 – 0.20 USD/km
\item Belgium, city zone, 3.5 – 12 tons: 0.12 – 0.22 USD/km
\end{itemize}

Finally, for the area-based scheme, as of June 2021, London's congestion charge zone covering 0.14 million residents and 20.7 sq km. charges 20.92 USD from 7AM to 10PM, 7 days a week. Given that our toll area is much larger and population density much smaller, the toll rates are considerably lower in comparison.

\subsection{Performance measures}\label{sec:Sec5_4}
In assessing the various congestion pricing policies, we examine several measures of performance. First, social welfare, second, network performance metrics, and third, travel/activity patterns and logistics operations indicators. These are discussed in turn. 

\subsubsection{Social welfare}
The total social welfare, denoted by $SW$ is given by, 
\begin{equation}
    SW = PS + CS - EC,
\end{equation}
where PS denotes the producer surplus, CS, the consumer surplus, and EC the emission costs. \\
\textit{Producer Surplus}: We assume that the producer in our context is the government or regulator, who operates the toll system and public transit and collects fuel taxes. The agency's revenue is the producer surplus and consists of three components: toll earning, public transit fare collection, and fuel tax revenue collected from passenger and freight vehicles. The toll earning is the toll revenue minus depreciation and operation costs, for which, we use a lower bound from various estimates -- 27 \% of the toll revenue \citep{kirk2017tolling}. \\
\textit{Consumer Surplus}: Both passenger travelers and freight agents are consumers. For passengers, we use a disaggregate utility-based accessibility measure originating from random utility theory. The accessibility measure of individual $n$, denoted $A_n$ equals the expected maximum utility $E(U_{an})$, which can be expressed as a logsum of the utilities over all alternative activity patterns and is given by,
\begin{equation}
A_n =  E(U_{an}) = \frac{1}{\mu} \textnormal{ln} \left[ \sum_{a \in C_n} exp(\mu V_{an}) \right],
\end{equation}
where, $V_{an}$ is the systematic component of utility $U_{an}$ of choosing an activity pattern $a$ from a set of alternative patterns $C_n$. The Activity-Based Accessibility measure (ABA) is obtained from the top of the hierarchical model structure (day pattern level, see Figure \ref{fig:SM_MT_Preday}) and is scaled with respect to cost and normalized to yield the consumer surplus:
\begin{equation}
ABA_n = \alpha_{nx}(A_n - A_n^0),
\end{equation}
where $A_n$ is the accessibility for individual $n$ under a specific policy, $A_n^0$ represents the accessibility under an alternative baseline scenario, and the scaling factor is given by \citep{dong2006moving}, 
\begin{equation}
\alpha_{nx} = \left( \frac{A_n^{\Delta x} - A_n}{\Delta x\sum_{j \in \mathcal{C}_n}p_{jn}t_j } \right)^{-1},
\end{equation}
where $A_n^{\Delta x}$ represents the accessibility under a scenario where the cost is increased by $\Delta x$ in all trips and activity schedules, $ \mathcal{C}_n$ represents the set of all possible activity schedules for individual $n$, $p_{jn}$ is the probability of individual $n$ choosing activity pattern $j$, and $t_j$ is the number of trips in activity pattern $j$. Note that we use the average per capita trip rate as an approximation of $\sum_{j \in \mathcal{C}_n}p_{jn}t_j$ for all individuals (given its complexity to calculate).  The passenger consumer surplus, denoted $CS_P$ is the sum of all travelers' ABA:
\begin{equation}
CS_P = \sum_n ABA_n
\end{equation}

On the freight side, our analysis is shipper-centric. In effect, we consider the freight shipper, carrier and receiver as one entity in the welfare analysis. The freight consumer surplus equals revenue minus cost. We assume that in the short run, the total revenue is not affected by congestion pricing, and thus, the change in surplus is solely determined by the change in logistics costs. On the one hand, this assumption implies that the commodity purchase and sales prices are not affected, which is reasonable since long-term decisions are out of our scope. On the other hand, we assume that B2C e-commerce demand changes more flexibly. We assume that shippers increase delivery fees to the tolled area and hence, the B2C commodity demand would decrease. The shipper’s revenue from B2C commodities decrease but it is compensated by the delivery fees collected, so the gross effect is unknown. We defer the detailed study of this to future research.

The daily TLC of a contract i belonging to shipper n is given as:

\begin{equation}
dailyTLC_i = \sum_{s \in i} (T_s + Y_s) + { D_i + I_i + K_i + Z_i \over period_i}  
\end{equation}

\begin{equation}
Transportation\: cost: T_s = c_s
\end{equation}

\begin{equation}
Capital\: cost\: during\: transportation: Y_s = d \cdot t_s \cdot v^{com_i} \cdot Q_i
\end{equation}

\begin{equation}
Deterioration\: and\: damage\: cost: D_i = d \cdot j \cdot g \cdot v^{com_i} \cdot Q_i
\end{equation}

\begin{equation}
Inventory\: holding\: cost: I_i = w^{com_i} \cdot q_i/2
\end{equation}

\begin{equation}
Capital\: cost\: of\: inventory : K_i = d \cdot v^{com_i} \cdot q_i/2 
\end{equation}

\begin{equation}
Stockout\: cost: Z_i = 
z \sqrt{LT_i \cdot \sigma_{Q_i}^2 + Q_i^2 \cdot \sigma_{LT_i}^2}
\end{equation}

To obtain an average weekday TLC, we simply divide the cost components $D_i$, $ I_i$, $K_i$,  $Z_i$ by the period of planning ${period}_i$. Table \ref{table:logisticscostpara} summaries the definitions and sources of parameters. The simulation outputs the daily transportation cost and travel time, which can be directly used to calculate the relevant cost components. 

\begin{table}[hbt!]
\renewcommand{\arraystretch}{0.85}
\centering
\caption{Parameters in the logistics cost function}\label{table:logisticscostpara}
    \resizebox{\textwidth}{!}{\begin{tabular}{p{0.1\linewidth}    p{0.3\linewidth}  p{0.6\linewidth} }
    \hline
        Notation & Description & Value \\ \hline
        $c_s$ & Transportation cost & The systematic component of the VOP model is used as the cost function. Distance, time, and toll values are obtained from the within-day simulation. \\ 
        $d$ & Discount rate of cash & 0.75\%/365 per day based on the 2021 secondary credit rate in Boston, Massachusetts. \\ 
        $t_s$ & Transportation time & Obtained from the within-day simulation. \\ 
        $v^{com_i}$ & Commodity-type-specific value per unit of weight & Estimated using the 2012 micro-sample of the U.S. Commodity Flow Survey. \\ 
        $Q_i$ & Total demand by weight & Simulated from commodity contract or e-commerce model; unit is kg per contract period. \\ 
        $j$ & Shrinkage rate & Assumed as 1\%. \\ 
        $g$ & Average period to receive an insurance claim & Assumed as 26.7 days. \\ 
        $w^{com_i}$ & Storage cost per unit of weight & Refer to  Equation \ref{eqn:WET}. \\ 
        $q_i$ & Shipment size of by weight & Simulated from shipment frequency \& size model or e-commerce model. \\ 
        $z$ & z-score & Assumed as 1.03 (85\% of cycles without shortage). \\ 
        $LT_i$ & Lead time & $LT_i = t_{s_i} + otherLT_i$. $otherLT_i$ are assumed as follows: 2 weekdays for B2C and 6 weekdays for B2B. \\ 
        $\sigma_{Q_i}^2$ & Variance of demand & Assumed 0 for B2B and obtained from the e-commerce simulation for B2C. \\
        $\sigma_{LT_i}^2$ & Variance of lead-time & Assumed two components are uncorrelated, then $\sigma_{LT_i}^2 = \sigma_{t_{s_i}}^2 +\sigma_{otherLT_i}^2 $. $\sigma_{t_{s_i}}^2$ is obtained from the within-day simulation. $\sigma_{otherLT_i}^2$ is assumed as 0.5.\\ \hline
\end{tabular}}
\end{table}

\textit{Emission Costs}: The last component of the total social welfare is emission cost – the social cost of CO2 emissions. It comprises a key negative externality. The conversion of CO2 emissions to cost is based on U.S. Environmental Protection Agency’s estimate for 2020, equivalent to 58 USD/ton of CO2 (U.S. EPA, 2016). For details on the emission model, we refer the reader to \cite{oke2020evaluating}.

\subsubsection{Network performance}
The measures of network performance we use are (1) Vehicle-Kilometres-Traveled (VKT) categorized by passenger mode and freight vehicle type, and (2) Travel Time Index (TTI), which is the ratio of congested to free-flow travel time, at the trip level. Congestion in a specific time period is quantified via a trip-length weighted TTI for all trips occurring in the time period.

\subsubsection{Travel/Activity patterns and logistics operations}
The indicators used to describe travel/activity patterns include number of trips by mode, number of activities by type, mode shares, trips patterns by time-of-day and trip length distributions. Indicators for logistic operations include shipment size and freight vehicle load factors (maximum shipment load in a tour divided by vehicle capacity, both measured by weight).

\subsubsection{Day-to-day learning and Stochasticity}
The performance measures for a given scenario are computed after five iterations of the day-to-day learning process, which ensures consistency between zone-to-zone travel private/public transit travel times and waiting times used in the pre-day and those obtained from the supply simulation (see Section \ref{sec:Supply}). The initial zone-to-zone travel times are already values from prior simulations and hence, five iterations suffice to attain a reasonable measure of consistency. Nevertheless, in the absence of computational constraints, performing a larger number of day-to-day iterations is recommended.  

It is also worth noting that the extent of stochasticity across multiple simulations of an 'average' day at 'convergence' is relatively low with regard to the key aggregate performance measures of interest:
\begin{itemize}
    \item Number of passenger trips: $0.07 \% - 0.2 \%$ 
    \item Number of freight trips: $0.1 \% - 4\%$
    \item Passenger VKT: $0.09 \% - 0.2\%$
    \item Freight VKT: $0.2 \% - 0.5\%$
\end{itemize}
The magnitude of differences across scenarios should be interpreted in the context of the above numbers. 

\section{Results and Discussion}\label{sec:X}
In this section, we report findings from the simulation experiments, focusing on overall welfare, distributional impacts, network performance, travel/activity patterns and logistics operations.

\begin{table}[h]
\renewcommand{\arraystretch}{0.85}
\centering
\caption{Number of trips and travelers tolled by entry time}\label{table:trips_tolled}
\small
\begin{tabular}{lllll}
\hline
                                                                & Entry time & Distance & Cordon & Area \\
                                                                \hline
\multirow{3}{*}{Passenger trips  (million)}     & AM            & 1.11     & 0.255  &      \\
                                                                & PM            & 1.42     & 0.181  &      \\
                                                                & All           & 2.57     & 0.436  & 3.78 \\
                                                                \hline
\multirow{3}{*}{Passengers (million)}          & AM            & 1.01     & 0.247  &      \\
                                                                & PM            & 1.17     & 0.202  &      \\
                                                                & All           & 1.82     & 0.447  & 2.42 \\
                                                                \hline
\multirow{3}{*}{Freight trips (thousand)}      & AM            & 75.3     & 13.1   &      \\
                                                                & PM            & 312      & 48.9   &      \\
                                                                & All           & 387      & 62.0   & 655  \\
                                                                \hline
\multirow{3}{*}{Freight vehicles (thousand)} & AM            & 30.1     & 8.73   &      \\
                                                                & PM            & 104      & 27.2   &      \\
                                                                & All           & 124      & 32.5   & 182 \\
                                                            \hline
\end{tabular}
\end{table}

Table \ref{table:trips_tolled} summaries the number of trips and the number of travelers tolled by trip’s entrance time, grouped into the AM peak (AM), PM peak (PM), and the entire day (ALL). Note that the toll period under Area is from 8 AM to 7 PM, which is longer than the other schemes (7:30 AM to 10:30 AM and 3:30 PM to 7:30 PM). For passengers, 52.6\% of the population are tolled under Area, 39.6\% of the population are tolled under Distance, whereas only 9.7\% of the population are tolled under Cordon. Similarly, among all carriers who have at least one shipment, a greater percentage is tolled under Area (81.4\%) than Distance (57.8\%), and significantly more than Cordon (15.9\%). We thus expect that the Cordon scheme would have a significantly smaller overall impact. Recall that we set upper bounds for the maximum daily toll for Distance policy – 10 USD for passenger car, 15 USD for LGV, 20 USD for HGV, and 25 USD for VHGV. It shows that 10.1\% of tolled passenger cars, 4.5\% of tolled LGVs, 13.4\% of tolled HGV, and 16.7\% of tolled VHGV reached this upper bound.

\begin{table}[!ht]
\renewcommand{\arraystretch}{0.85}
\caption{Average tolls paid (USD)}\label{table:tolls_paid}
\centering
\begin{tabular}{llll}
\hline
                    & Distance & Cordon & Area \\
                    \hline
Per passenger       & 3.31     & 2.79   & 2.65 \\
Per LGV             & 8.56     & 5.42   & 4.00 \\
Per HGV             & 11.52    & 6.15   & 6.00 \\
Per VHGV            & 15.30    & 7.83   & 7.50 \\
Per freight carrier & 19.54    & 11.84  & 9.35 \\
\hline
\end{tabular}
\end{table}

Among the travelers who pay tolls, the average tolls paid are summarized in Table \ref{table:tolls_paid}. The average toll per passenger, per freight vehicle, as well as per freight carrier are highest for the Distance-based scheme followed by the Cordon-based and Area schemes, respectively. The total toll revenue on an average weekday is 7.53 million USD under Distance, 1.56 million USD under Cordon, and 7.22 million USD under Area. To put these numbers in context, the realized toll revenues on an average day observed under the Stockholm congestion charging scheme in 2006 was around 0.68 million USD\footnote{adjusted to 2022 USD amounts; https://www.bls.gov/cpi/} \citep{eliasson2009stockholm} (this increased two-fold in 2016, see \cite{borjesson2018swedish}). The corresponding figures for Singapore in 2015 were approximately 0.59 million USD\footnote{https://ops.fhwa.dot.gov/publications/fhwahop08047/02summ.htm; https://mothership.sg/2018/04/erp-history-20-years/, Accessed 23 August, 2022} (2022 USD) and annual revenues from the London congestion charging scheme in 2018 were around 300 million USD\footnote{https://content.tfl.gov.uk/tfl-annual-report-and-statement-of-accounts-2018-19.pdf; Retreived 23 August, 2022}.  

\subsection{Welfare}
Table \ref{table:Welfare} summarizes the change in total social welfare compared against the base scenario for an average weekday, expressed in million USD. Clearly, the congestion pricing policies reduce the use of private vehicles and encourage public transit ridership, thus improving public transport revenue and reducing fuel tax revenue. We can also observe that in fact the toll revenues are significantly higher than the changes in consumer surplus or net user benefits (for both passenger and freight agents), for all the three pricing schemes. This finding is consistent with the literature, for example, \cite{eliasson2006equity} quantified the net total benefits to revenue ratio (assuming toll revenues are refunded uniformly through a lump sum allocation) from the Stockholm congestion charging scheme to be 0.32, while \cite{de2005congestion} reported a net benefit to total revenue ratio of 0.28 for a cordon scheme.  The ratio of change in social welfare to toll revenues are 0.30 (2.27/7.53), 0.11 (0.167/1.56) and 0.238 (1.72/7.22) for the distance, cordon and area schemes, respectively. Overall, all three pricing policies improve overall welfare. 

\begin{table}[!ht]
\renewcommand{\arraystretch}{1}
\caption{Welfare gains under different scenarios relative to baseline (USD)}\label{table:Welfare}
\small
\hspace{-0 cm}
\begin{tabular}{lllll}
\hline
           \multirow{2}{*}{Category }                  &     \multirow{2}{*}{ Component (M USD / weekday) }                  & \multicolumn{3}{c}{Scenario}            \\
                                   &    & Distance & Cordon  & Area    \\
\hline
\multirow{3}{*}{Change in producer surplus} & Toll earning                      & +5.50    & +1.14   & +5.27   \\
                                            & Public transport revenue    & +0.0545  & +0.0124 & +0.0586 \\
                                            & Fuel tax revenue             & -0.0864  & -0.0152 & -0.0872 \\
                                            \hline
\multirow{2}{*}{Change in consumer surplus} & Passengers                   & -3.44    & -1.01  & -3.73   \\
                                            & Freight shippers/carriers   & -0.167   & -0.0529 & -0.228  \\
                                            \hline
Change in emission costs              & CO2 emissions                & -0.412   & -0.0932 & -0.445  \\
\hline
\multicolumn{2}{l}{Change in total social welfare}                      & +2.27    & +0.167   & +1.72  \\
\hline
\end{tabular}
\end{table}



In terms of the relative performance of the three pricing schemes, the distance-based scheme outperforms the other two (gross social welfare of 2.27 million USD versus 1.72 and 0.167 for the area and cordon schemes). The distance-based scheme clearly internalizes congestion externalities more effectively by charging higher for longer trips that contribute more to congestion. This finding of the superiority of distance-based scheme relative to access- and area-based schemes has been reported for stylized networks and bottleneck models in the literature \citep{lehe2017downtown,liu2022managing}. \cite{lentzakis2020hierarchical} also arrive at similar conclusions, estimating significantly higher welfare gains for a distance-based scheme relative to a cordon scheme. Comparing the distance-based and area-based schemes, we find that although the gross welfare improvements are relatively close, the reduction in consumer surplus is smaller under Distance (passengers: -3.44 million USD, freight: -0.167 million USD) compared with Area (passengers: -3.73 million USD, freight: -0.228 million USD). Finally, we note that part of the reason for the poor performance of the cordon scheme may be its design; it is plausible that higher benefits may be attained if there is a second cordon is closer to downtown Boston.

A third observation that we make is that the overall net benefits (prior to any refunding or use of the toll revenues) are negative for both passenger and freight agents, implying that the overall value of gains in travel time are smaller than the total tolls paid. This finding is also consistent with the literature, see for example, \cite{eliasson2006equity} and \cite{de2005congestion}. Nevertheless, it should be pointed out that in the case of passenger travel, the extent and type of heterogeneity can determine whether net user benefits (prior to the use of toll revenues) are positive or negative. For instance, \cite{van2011winning} show that congestion pricing can leave a majority of travellers better off even without redistribution of toll revenues. They employ Vickrey's dynamic model of bottleneck congestion with continuously distributed values of time and schedule delay. Using a similar bottleneck model that also includes a transit alternative (mode and departure time choice described by a logit model with heterogeneity in value of time, schedule delay, early and late), \cite{CHEN2023104121} find that when the coefficient of variation in the value of is large (exceeds around 0.5 in their experiments), the net user benefits start to become positive (even before accounting for the use of toll revenues). Similarly, \cite{he2021validated}  and \cite{lentzakis2023predictive} find positive net benefits even prior to any toll revenue redistribution using large-scale simulation models. 

Finally, in terms of emission costs, the area and distance-based schemes again yield significantly larger benefits (reduction of 0.445 and 0.412 million USD per weekday) compared to the cordon scheme. 

\subsection{Distributional Impacts - Passenger}
Distributional impacts have long been a focal point of political opposition to congestion pricing. Pricing may viewed as being `unfair' if it is seen as disproportionately benefiting high income groups and burdening low income groups \citep{eliasson2016congestion}. As noted in Section one, the evidence on this is mixed with some studies suggesting that congestion pricing is regressive \citep{evans1992road,arnott1994welfare,seshadri2022congestion}, whereas others indicating it is progressive \citep{eliasson2006equity,santos2004distributional}. Ultimately, the question of distributional impacts cannot be resolved theoretically and is a matter for empirical investigation. 
Our analysis of distributional outcomes implicitly assumes that value of time is correlated with income and that there is a one-to-one relationship between VOT and income. \cite{verhoef2004product} have cautioned against viewing the VOT distribution as simply representing the income distribution since empirical studies have identified other correlates of income (see also \cite{lehe2020winners} on this). Nevertheless, due to the lack of sufficient data on this, as is standard practice, we assume that VOT is correlated with income.  

\begin{table}[!h]
\caption{Distributional impacts - Passenger}\label{table:Dist_Pass}
\renewcommand{\arraystretch}{0.85}
\footnotesize
\hspace{-1.5 cm}
\begin{tabular}{llllll}
\hline
Profiles                                                                              & All              & Group 1         & Group 2     & Group 3    & Group 4         \\
\hline
\multicolumn{6}{l}{Distance}                                                                                                                                            \\
Range of daily   surplus (USD/weekday)                                                & {[}-13.1,9.2{]} & {[}-13.1,-1.3) & {[}-1.3,0) & {[}0,1.3) & {[}1.3,9.2{]} \\
Percentage in   population                                                            & 100\%            & 22.8\%          & 37.1\%      & 33.3\%     & 6.8\%           \\
Avg daily   surplus (USD/weekday)                                                     & -0.84           & -4.73           & -0.33      & 0.28      & 4.1            \\
Avg number of car trips with either OD in toll area during toll periods             & 0.554            & 0.725           & 0.561       & 0.454      & 0.431           \\
Avg distance traveled by car within toll area during toll periods (km)              & 5.19             & 7.27            & 5.24        & 4.00       & 3.82            \\
Proportion of households with work/education location in toll area & 0.683            & 0.933           & 0.623       & 0.524      & 0.953           \\
Proportion of work   activities among all daily activities                            & 0.272            & 0.238           & 0.257       & 0.282      & 0.420           \\
Avg household income   (thousand \$/year)                                             & 86.31            & 79.24           & 86.51       & 87.15      & 105            \\
\hline

\multicolumn{6}{l}{Cordon}                                                                                                                                              \\
Range   of daily surplus (USD/weekday)                                                & {[}-14.7,3.75{]} & {[}-14.7,-1.3) & {[}-1.3,0) & {[}0,1.3) & {[}1.3,3.75{]} \\
Percentage in population                                                              & 100\%            & 8.1\%           & 34.2\%      & 51.3\%     & 6.3\%           \\
Avg daily surplus (USD/weekday)                                                       & -0.25          & -6.0           & -0.28      & 0.41      & 2.0           \\
Avg number of car trips with either OD in toll area   during toll periods             & 0.561            & 1.16            & 0.425       & 0.397      & 1.87            \\
Avg number of car trips entering toll area during   toll periods                      & 0.0953           & 1.13            & 0.0108      & 0.000100   & 0.000179        \\
Avg distance traveled by car within toll area during   toll periods (km)              & 5.50             & 11.1            & 3.38        & 4.37       & 19.1            \\
Proportion of households with work/education location in toll area & 0.683            & 1.000           & 0.613       & 0.648      & 0.950           \\
Proportion of work activities among all daily   activities                            & 0.272            & 0.181           & 0.262       & 0.276      & 0.416           \\
Avg household income (thousand \$/year)                                               & 86.3             & 78.0            & 84.5        & 86.7       & 105             \\
\hline
\multicolumn{6}{l}{Area}                                                                                                                                                \\
Range of daily surplus (USD/weekday)                                                  & {[}-6.9,10.1{]} & {[}-6.9,-1.3) & {[}-1.3,0) & {[}0,1.3) & {[}1.3,10.1{]} \\
Percentage in population                                                              & 100\%            & 31.5\%          & 27.9\%      & 28.9\%     & 11.7\%          \\
Avg daily surplus change   (USD/weekday)                                              & -0.93           & -4.34           & -0.33      & 0.34      & 3.63            \\
Avg number of car trips   with either OD in toll area during toll periods             & 0.820            & 1.16            & 0.304       & 0.348      & 2.30            \\
Avg distance traveled by   car within toll area during toll periods (km)              & 6.76             & 10.7            & 1.69        & 1.74       & 20.6            \\
Proportion of households with work/education location in toll area & 0.683            & 0.853           & 0.551       & 0.551      & 0.865           \\
Proportion of work   activities among all daily activities                            & 0.272            & 0.254           & 0.270       & 0.268      & 0.334           \\
Avg household income   (thousand \$/year)                                             & 86.3             & 82.5            & 86.3        & 88.4       & 91.5       \\
\hline    
\end{tabular}
\end{table}


 \begin{figure}[!h]
\hspace{-1 cm}
\includegraphics[scale=0.6]{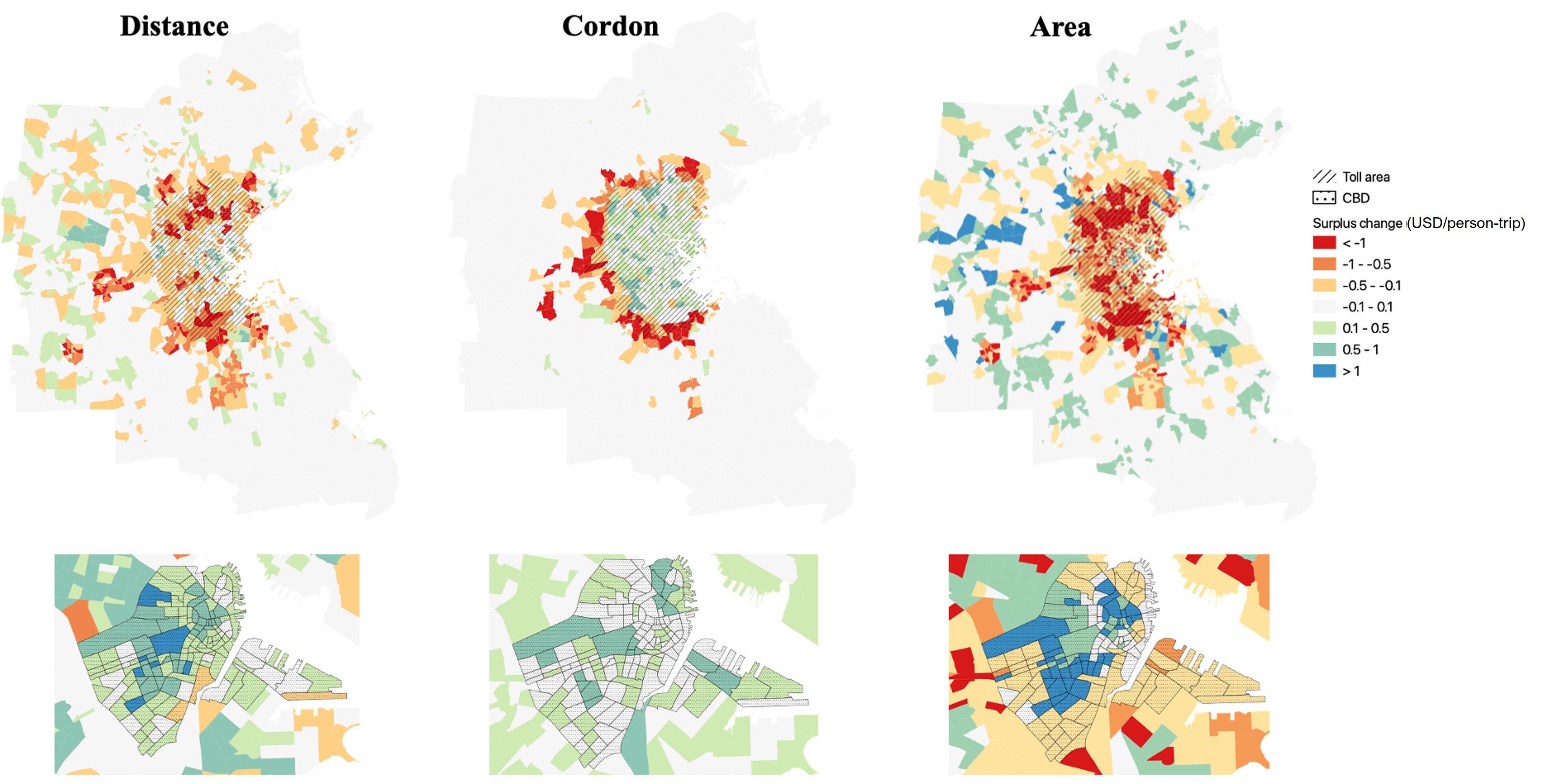}
  \caption{Spatial distribution of passenger surplus (by household location)}  \label{fig:Spatial_Pass}
\end{figure}

With microsimulation, we can compute each individual’s change in daily surplus and thus, identify “winners” and “losers” under congestion pricing by describing profiles of passengers belonging to groups with different ranges of daily surplus change. The definition of these groups is based on the assumption that a daily change in user benefits (surplus) of $\pm 0.50$ USD per trip can be considered small (with an average trip rate of 2.6, this implies a daily surplus change within $\pm 1.30$ USD per day). The grouping is instructive in analyzing distributional outcomes and identifying the differences between these sets of individuals. Under the distance-based scheme (Table \ref{table:Dist_Pass}), group 1 loses between 13.1 to 1.3 USD/person/weekday and group 4 gains from 1.3 to 9.2 USD/person/weekday while group 2 and group 3 are relatively less affected with changes of no more than 1.30 USD/person/weekday. Group 1 consists of 22.8\% of the population, group 2 consists of 37.1\%, group 3 consists of 33.3\%, and group 4 is the minority, comprising 6.8\%. Prior to the use of toll revenues, overall, 40.0\% of the population benefit from the distance-based scheme while 60.0\% are worse off. 

Compared to the entire population, both group 1 (largest losses) and 4 (largest gains) have a high proportion of individuals with a fixed household work or education location within the toll area (93\% and 95\%, respectively). Interestingly, Group 4 have higher household income but make fewer and shorter trips on average (resulting in relatively lower tolls paid) and gain the most from the pricing scheme. As can be seen from Figure \ref{fig:Spatial_Pass}, they also tend to reside close to the financial district in downtown Boston (lower panel in Figure \ref{fig:Spatial_Pass}). In contrast, group 1 are more likely to stay outside the toll area or at the periphery, make more trips on average within the tolled area and have lower household incomes. In this regard, despite yielding the largest welfare gains, the current design of the distance-based scheme does appear to be regressive, as can be seen from the trend of increasing average incomes from group 1 to group 4. A lump-sum redistribution of toll revenues, while improving equity, will not guarantee Pareto improvement, nor will it make the scheme progressive. On the other hand, other avenues for use of the toll revenues, such as increasing public spending (to improve public transport or reduce car costs) or decreasing taxes will alter the distribution of benefits and have the potential to make the scheme progressive (see for example, \cite{eliasson2006equity}). Alternatively, other designs of the tariff scheme such as a two-part tariff may redress some of the inequities in the distribution of benefits given the large gains observed for individuals with short trips and high incomes. Note that although the increasing trend in consumer surplus is associated with an increasing average income across the four groups, when considering individual travelers, the consumer surplus is not monotonic with the value of time.  The non-monotonicity of consumer surplus (losses or gains) with respect to value of time has been reported for the bottleneck model \citep{van2011winning} and a static model with a mixed-bus alternative \citep{lehe2020winners}. 

It is also noteworthy that group 4 has a higher proportion of work activities than group 1 and these are associated with higher values of time. Group 2 and group 3 are more homogeneous compared to group 1 or group 4 and their profiles are close to population average. 


 
Under the Cordon scheme, Group 1 loses 14.7 to 1.3 USD/person/weekday, group 4 gains 1.3 to 3.75 USD/person/weekday, and group 2 and group 3 have loss or gain of no more than 1.30 USD/person/weekday. Group 1 consists of 8.1\% of the population, group 2 consists of 34.2\%, and group 3 and group consist of 51.3\% and 6.3\% respectively. Overall, 57.7\% of the population gain and 42.3\% lose. Compared to the distance-based scheme, the proportion of big `losers' (group 1) is much smaller due to the significantly smaller proportion of trips tolled. Concurrently, the proportion the large `gainers' is also much smaller due to the significantly smaller travel time gains. Once again, as expected, both group 1 and group 4 are more likely to have a fixed household/work/education location in the toll area. Unlike the distance-based scheme, group 4 has a significantly large number of trips within the tolled area and these are almost exclusively internal trips that are not tolled but benefit from the travel time savings. In contrast, group 1 makes a higher number of entering trips (as expected) than the population average, thus incurring the toll cost. The cordon scheme also appears to be regressive, with an increase in average household income observed from group 1 to group 4. The large  `gainers' are high income households with an average annual income of around \$105,000.     

Under the area-based scheme, Group 1 consists of 31.5\% of the population and loses 1.30 to 6.9 USD/person/weekday; group 2 consists of 27.9\% and losses no more than 1.30 USD/person/weekday; group 3 consists of 27.9\% and gains no more than 1.30 USD/person/weekday; group 4 takes up 11.7\% and gains 1.30 to 10.1 USD/person/weekday. The proportions of large `losers' and `winners' are much higher than the other two scenarios. In the area-based scheme users pay a flat daily fee to travel within the tolled area, and thus, the individuals who gain the most tend to have a significantly higher number of trips in the tolled area than average. These are not necessarily those with the highest incomes living in the vicinity of downtown Boston, and thus the average incomes of the four groups are closer to each other than the other schemes. This can also be seen in Figure \ref{fig:Spatial_Pass}, which shows that while the individuals who lose the most under the distance-based scheme tend to reside in the outer parts of the tolled region or outside the periphery, in the area-based scheme they tend to be distributed throughout the tolled region. Thus, the area-based scheme appears to be the least regressive amongst the three schemes. However, clearly, in terms of overall welfare gains and travel time savings, it is not as effective as the distance-based scheme which more directly charges for the externalities of travel.   

\subsection{Distributional Impacts - Freight}
We perform a similar analysis for freight by describing profiles of shippers belonging to groups with different ranges of daily surplus change. As in the analysis of passengers, we divide shippers into four groups depending on their change in daily surplus under the different tolling schemes.

\begin{table}[!ht]
\caption{Distributional impacts - Freight}\label{table:Dist_Freight}
\renewcommand{\arraystretch}{0.85}
\footnotesize
\hspace{-2 cm}
\begin{tabular}{llllll}
\hline  
Profiles                                                                       & All             & Group 1          & Group 2     & Group 3    & Group 4         \\
\hline  
\multicolumn{6}{l}{Distance}                                                                                                                                     \\
\hline  
Range of daily   surplus change (USD/weekday)                                  & {[}-322,161{]}  & {[}-322,-20.0)   & {[}-20.0,0) & {[}0,20.0) & {[}20.0,161{]}  \\
Proportion in   shipper establishments                                         & 100\%           & 9.1\%            & 50.0\%      & 36.7\%     & 4.2\%           \\
Avg daily   surplus change (USD/weekday)                                       & -3.85           & -61.3            & -4.59       & 3.28       & 66.3            \\
Avg number of   trips with either OD in toll area during toll periods          & 8.92            & 17.2             & 7.15        & 7.28       & 26.5            \\
Avg distance   traveled within toll area during toll periods (km)              & 64.2            & 154              & 48.3        & 51.3       & 174             \\
Avg number of shipments   with either OD in toll area                          & 8.05            & 15.8             & 6.45        & 6.82       & 21.2            \\
Avg shipment value by   weight (\$/kg)                                         & 11.4            & 8.76             & 10.3        & 12.8       & 16.8            \\
Proportion of having   establishment or pickup/delivery stops within toll area & 0.664           & 0.992            & 0.615       & 0.613      & 0.993           \\
Avg employment size                                                            & 18.0            & 3.21             & 17.7        & 17.9       & 54.0            \\
\hline  
\multicolumn{6}{l}{Cordon}                                                                                                                                       \\
\hline  
Range of daily surplus   change (USD/weekday)                                  & {[}-351,87.2{]} & {[}-351,-20.0)   & {[}-20.0,0) & {[}0,20.0) & {[}20.0,87.2{]} \\
Proportion in shipper   establishments                                         & 100\%           & 4.4\%            & 39.4\%      & 53.0\%     & 3.2\%           \\
Avg daily surplus change   (USD/weekday)                                       & -1.22           & -86.2            & -3.94       & 5.15       & 44.4            \\
Avg number of trips with   either OD in toll area during toll periods          & 9.10            & 24.9             & 7.32        & 7.72       & 32.1            \\
Avg number of trips   entering toll area during toll periods                   & 1.43            & 17.3             & 0.730       & 0.441      & 4.58            \\
Avg distance traveled   within toll area during toll periods (km)              & 68.8            & 184              & 56.4        & 59.7       & 214             \\
Avg number of shipments   with either OD in toll area                          & 8.05            & 18.9             & 6.54        & 6.72       & 33.7            \\
Avg shipment value by   weight (\$/kg)                                         & 11.4            & 6.42             & 9.80        & 12.5       & 18.4            \\
Proportion of having   establishment or pickup/delivery stops within toll area & 0.664           & 0.997            & 0.665       & 0.616      & 0.990           \\
Avg employment size                                                            & 18.0            & 1.71             & 16.5        & 17.8       & 61.5            \\
\hline  
\multicolumn{6}{l}{Area}                                                                                                                                         \\
\hline  
Range of daily surplus   change (USD/weekday)                                  & {[}-223,217{]}  & {[}-223,-20.0{]} & {[}-20.0,0) & {[}0,20.0) & {[}20.0,217{]}  \\
Proportion in shipper   establishments                                         & 100\%           & 20.4\%           & 29.7\%      & 34.8\%     & 15.2\%          \\
Avg daily surplus change (USD/weekday)                                         & -5.26           & -75.2            & -3.68       & 4.23       & 63.8            \\
Avg number of trips with   either OD in toll area during toll periods          & 15.1            & 22.5             & 8.43        & 10.5       & 28.6            \\
Avg distance traveled   within toll area during toll periods (km)              & 116             & 133              & 69.1        & 78.4       & 268             \\
Avg number of shipments   with either OD in toll area                          & 8.05            & 11.7             & 5.53        & 5.66       & 13.5            \\
Avg shipment value by   weight (\$/kg)                                         & 11.3            & 10.1             & 9.43        & 12.7       & 13.5            \\
Proportion of having   establishment or pickup/delivery stops within toll area & 0.664           & 0.990            & 0.482       & 0.484      & 0.990           \\
Avg employment size                                                            & 18.0            & 12.3             & 17.6        & 18.2       & 25.8           \\
\hline  
\end{tabular}
\end{table}

Under the distance-based scheme, group 1 comprises 9.1\% of all shippers and loses from 322 to 20 USD/shipper/weekday; group 2 comprises 50.0\% and loses no more than 20 USD/shipper/weekday; group 3 makes up 36.7\% and gains no more than 20 USD/shipper/weekday; group 4 comprises a small proportion – 4.2\% and gains from 20 to 161 USD/shipper/weekday. A daily surplus change of 20 USD is assumed to be small given that it constitutes 10\% of the median daily total logistics cost of a shipper under the base scenario. The overall percentage of ´winners' and `losers' is remarkably similar to the passenger case, with 40.9\% of shippers gaining and 59.1\% of shippers losing due to the pricing scheme. Although both group 1 and group 4 have a higher average number of shipments with an origin or destination in the tolled area (15.8 and 21.4, respectively, compared to the population average of 8.05), the average shipment value by weight of group 4 (16.8 \$ per kg) is significantly higher than that of group 1 (8.76 \$ per kg). Since the total logistics cost is positively correlated with shipment value by weight, given the same travel time savings, shippers with higher shipment values benefit more. Further, we also observe that group 1 has a smaller average employment size whereas group 4 has a significantly larger size, indicating that smaller business establishments tend to lose benefits. Thus, as in the case of passengers, the distance-based scheme also appears to be regressive with regard to distributional impacts on shippers. 

  \begin{figure}[!h]
\hspace{-1.5 cm}
\includegraphics[scale=0.6]{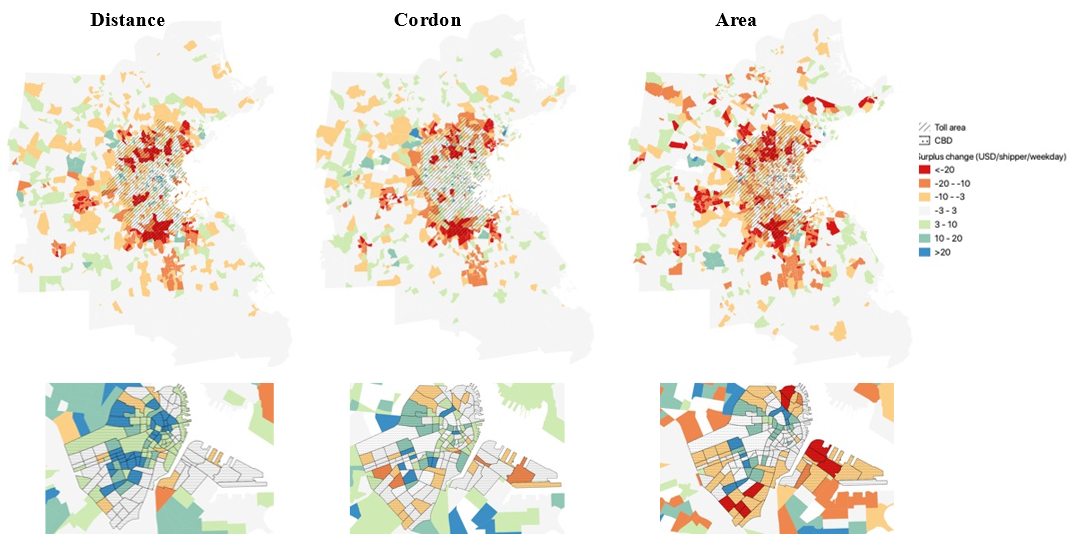}
  \caption{Spatial distribution of freight surplus (by shipper location)}  \label{fig:Spatial_freight}
\end{figure}

Under the Cordon scheme, Group 1 constitutes 4.4\% of the shipper population who loses from 20 to 351 USD/shipper/weekday; group 2 comprises 39.4\% and loses no more than 20 USD/shipper/weekday; group 3 makes up 53.0\% and gains no more than 20 USD/shipper/weekday; group 4 constitutes 3.2\% and gains 20 to 87.2 USD/shipper/weekday. Overall, 56.2\% of shippers gain and 43.8\% lose. As with the distance-based scheme, both group 1 and group 4 have a significantly higher number of trips with either the origin or destination in the tolled region. However, in case of group 4, a large percentage of these trips are internal trips that are not tolled whereas group 1 has a large percentage of trips that cross the cordon. Once again, the scheme appears to be regressive since shippers that benefit the most tend to be large establishments with a high shipment value.  

Finally, under the area-based scheme, the profiles of the four groups are as follows: Group 1 comprises 20.4\% of the shipper population and loses 20 to 223 USD/shipper/weekday; group 2 makes up 29.7\% of the population and loses no more than 20 USD/shipper/weekday; group 3 makes up 34.8\% and gains no more than 20 USD/shipper/weekday; group 4 constitutes 15.2\% and gains 20 to 217 USD/shipper/weekday. 50.0\% of shippers gain and 50.0\% lose. The profiles of the four groups suggest that the area-based scheme is less regressive than the distance and cordon schemes with the variation in average employment size and average shipment value being small across the four groups. 

Figure \ref{fig:Spatial_freight} shows the average daily surplus change of shippers by shipper business establishment location. Observe that there are differences in the spatial patterns of daily surplus changes for freight agents and passengers. One explanation is that the shipper’s establishment location is not necessarily a frequently visited stop, especially if the shipper does not use in-house fleets. This is also the reason why there can be establishments within the toll cordon under the cordon-based scheme that incur a negative surplus. A second explanation for the differences relative to the passenger case is the great variability in shippers’ logistics operations, such as varying numbers of shipments, shipment values, fleet sizes, etc.

\subsection{Network Performance}
The impacts of the tolling schemes on VKT are summarized in Table \ref{table:VKT}. The distance and area-based schemes both yield significant reductions in passenger car VKT overall (10\% and 5\%, respectively for internal trips and 5.2\% and 2.6\%, respectively for connection trips). As expected, the reductions in VKT in the distance-based scheme come  solely from the peak periods (when the tolls are in effect) whereas for the area-based scheme they are distributed throughout the entire day. In both schemes, the reductions are larger for internal trips than connection
trips, potentially due to a larger share of discretionary activities (which may be cancelled or performed at a location closer to home) in internal trips compared to connection trips. In contrast, the cordon-based scheme is far less effective, resulting in an overall VKT reduction of only 0.7\% for connection trips, while in fact increasing the VKT of internal trips by 1.5\%. This is the result of induced demand and destination changes of internal trips (which are not tolled) due to the reduced congestion within the cordon. Nevertheless, we once again caution that different cordon designs may result in drastically different outcomes. In this case, it is evident that a second cordon closer to downtown Boston may be necessary to more substantially improve network performance.  

The impacts on freight VKT are more complex, being the combined result of changes in shipment size, ecommerce demand, vehicle operations planning and route choice. The relative reductions in freight VKT are once again the largest for the distance-based scheme for both internal and connection trips. Interestingly, internal the VKT of internal freight trips shows a reduction (although small) even under the cordon scheme despite the reduction of network congestion. The three schemes show differing impacts on VKT by vehicle type (LGV, HGV, VHGV).

\begin{table}[ht!]
\caption{VKT by vehicle type, OD pair, and departure time}\label{table:VKT}

\renewcommand{\arraystretch}{1}
\small
\hspace{-0.5 cm}
\begin{tabular}{l|l|llll|llll}
\hline
\multirow{2}{*}{Time-of-Day} & \multirow{2}{*}{Vehicle Type} & \multicolumn{4}{c|}{Internal Trips}   & \multicolumn{4}{c}{Connection Trips} \\
                             &                               & Base\textsuperscript{*}  & Distance & Cordon  & Area    & Base\textsuperscript{*}  & Distance & Cordon  & Area    \\
                             \hline
\multirow{4}{*}{All Day}     & Cars                          & 27.03 & -10.00\% & 1.50\%  & -5.00\% & 50.43 & -5.20\%  & -0.70\% & -2.60\% \\
                             & LGV                           & 1.46  & -9.00\%  & -0.90\% & -6.50\% & 2.49  & -5.50\%  & -3.80\% & -3.00\% \\
                             & HGV                           & 0.16  & -11.20\% & -0.40\% & -7.90\% & 0.4   & -6.00\%  & -3.90\% & -3.00\% \\
                             & VHGV                          & 0.08  & -10.60\% & -2.10\% & -5.50\% & 0.45  & -6.90\%  & -2.60\% & -2.90\% \\
                             \hline
\multirow{4}{*}{AM Peak}     & Cars                          & 4.68  & -14.40\% & 2.40\%  & -3.80\% & 10.3  & -8.30\%  & -3.10\% & -1.70\% \\
                             & LGV                           & 0.16  & -10.50\% & -0.90\% & -2.70\% & 0.34  & -6.70\%  & -5.40\% & -0.60\% \\
                             & HGV                           & 0.02  & -13.40\% & -1.70\% & -5.00\% & 0.05  & -6.60\%  & -4.70\% & -0.50\% \\
                             & VHGV                          & 0.01  & -11.20\% & -1.00\% & -4.20\% & 0.06  & -8.20\%  & -5.00\% & -1.80\% \\
                             \hline
\multirow{4}{*}{PM Peak}     & Cars                          & 6.22  & -13.20\% & 1.90\%  & -4.10\% & 11.6  & -8.00\%  & -1.40\% & -1.90\% \\
                             & LGV                           & 0.64  & -11.00\% & -1.30\% & -4.10\% & 1.14  & -6.90\%  & -5.50\% & -1.70\% \\
                             & HGV                           & 0.07  & -12.60\% & -0.70\% & -5.60\% & 0.17  & -8.10\%  & -6.40\% & -1.90\% \\
                             & VHGV                          & 0.04  & -14.60\% & -0.70\% & -7.20\% & 0.2   & -7.70\%  & -7.40\% & -3.40\% \\
                             \hline
\multicolumn{10}{l}{ \textsuperscript{*} Units: million kilometres} 
\end{tabular}
\end{table}

In order to examine travel time improvements due to the pricing policies, we examine trips to, from and within the more congested central business district (CBD) within the toll area. The distance-weighted TTI for internal and connection trips to the CBD are summarized in Table \ref{table:TTI}. The distance-based scheme yields the highest reductions in TTI for internal trips (13.1\% and 12.7\% in the AM and PM peak, respectively) whereas the reductions in the area-based scheme are more modest (7.6\% and 8.5\%, respectively). The cordon-based schemes yields the smallest reductions in TTI (3.1\% and 2.6\% in the AM and PM peak, respectively). The trends for connection trips are similar. 

\begin{table}[]
\renewcommand{\arraystretch}{0.85}
\small
\caption{Travel Time Index (TTI) for CBD trips}\label{table:TTI}
\begin{tabular}{l|llll|llll}
\hline
\multirow{2}{*}{Period} & \multicolumn{4}{c|}{Internal Trips (CBD)}                                                                                                                                                              & \multicolumn{4}{c}{Connection Trips (CBD)}                                                                                                                                                           \\
                        & Base & Distance                                                       & Cordon                                                        & Area                                                          & Base & Distance                                                      & Cordon                                                        & Area                                                          \\
                        \hline
AM Peak                 & 1.33 & \begin{tabular}[c]{@{}l@{}}1.15\\    \\ (-13.1\%)\end{tabular} & \begin{tabular}[c]{@{}l@{}}1.28\\    \\ (-3.1\%)\end{tabular} & \begin{tabular}[c]{@{}l@{}}1.22\\    \\ (-7.6\%)\end{tabular} & 1.27 & \begin{tabular}[c]{@{}l@{}}1.15\\    \\ (-9.4\%)\end{tabular} & \begin{tabular}[c]{@{}l@{}}1.20\\    \\ (-5.5\%)\end{tabular} & \begin{tabular}[c]{@{}l@{}}1.17\\    \\ (-7.9\%)\end{tabular} \\
\hline
PM Peak                 & 1.28 & \begin{tabular}[c]{@{}l@{}}1.12\\    \\ (-12.7\%)\end{tabular} & \begin{tabular}[c]{@{}l@{}}1.25\\    \\ (-2.6\%)\end{tabular} & \begin{tabular}[c]{@{}l@{}}1.18\\    \\ (-8.5\%)\end{tabular} & 1.21 & \begin{tabular}[c]{@{}l@{}}1.11\\    \\ (-8.1\%)\end{tabular} & \begin{tabular}[c]{@{}l@{}}1.16\\    \\ (-4.0\%)\end{tabular} & \begin{tabular}[c]{@{}l@{}}1.11\\    \\ (-8.7\%) 

\end{tabular} \\
\hline
\end{tabular}
\end{table}

\subsection{Travel and Activity Patterns}
In this section, we examine the impacts of the three pricing schemes on activity generation (passenger and freight), mode shares and departure time patterns.

  \begin{figure}[!h]
\hspace{-1 cm}
\includegraphics[scale=0.67]{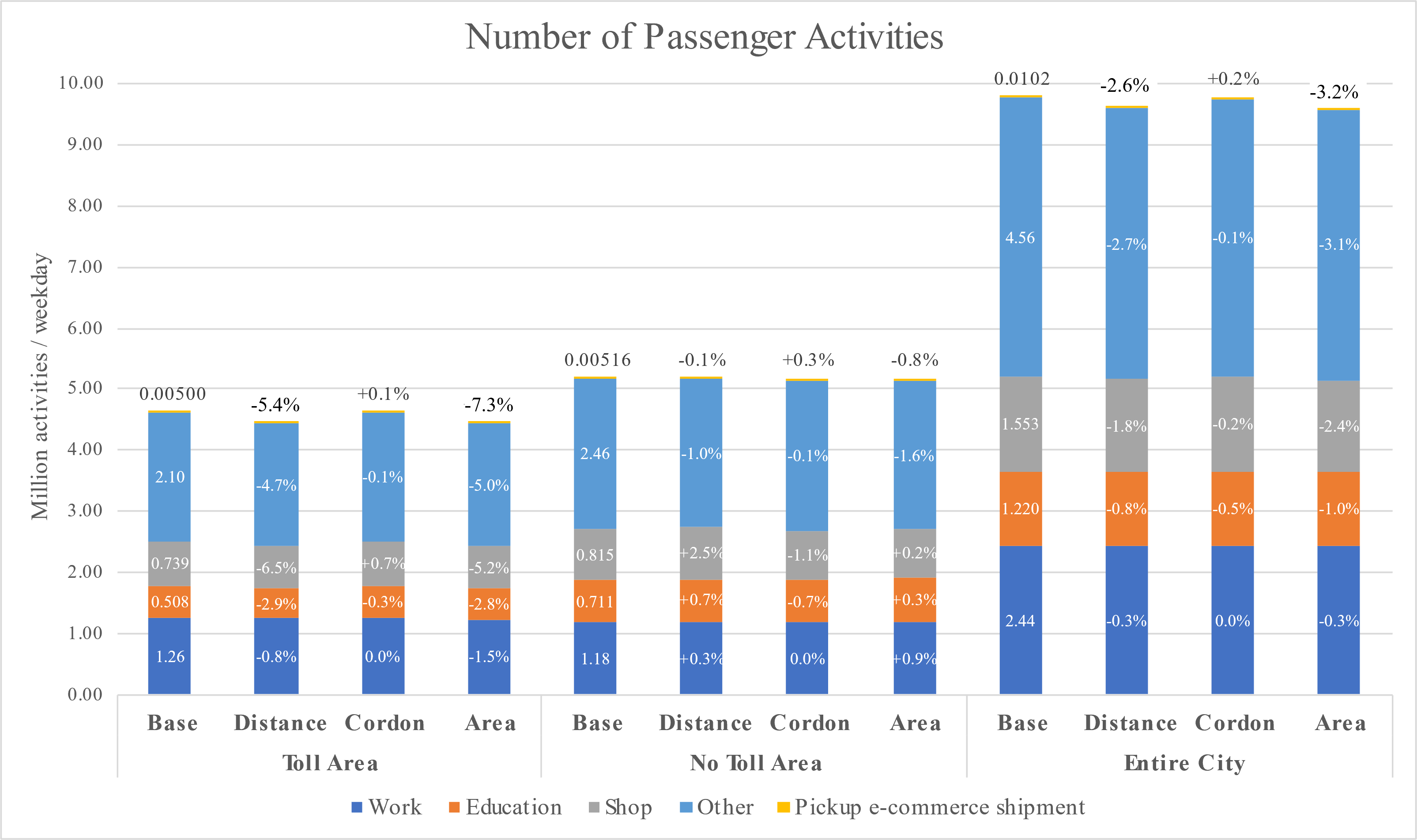}
  \caption{Impacts on activity generation}  \label{fig:Act_gen}
\end{figure}

\subsubsection{Activity Generation}
As expected, under both the distance and area-based schemes, a significant reduction in discretionary activities (for example, 6.5\% and 5.2\% reduction in shopping activities, respectively) is observed in the tolled area (see Figure  \ref{fig:Act_gen}). Work and education activities, which are typically less elastic are affected to a smaller extent. The impacts on activities are likely to have welfare effects (for example, on economic productivity and the labor market) that are not quantified in our model. Observe also that the design of the pricing scheme can impact the destination choices of shopping activities; for instance, under the distance-based scheme, we see a small increase in shopping activities (+0.7\%) outside the tolled area.

  \begin{figure}[!h]
\hspace{-1 cm}
\includegraphics[scale=0.67]{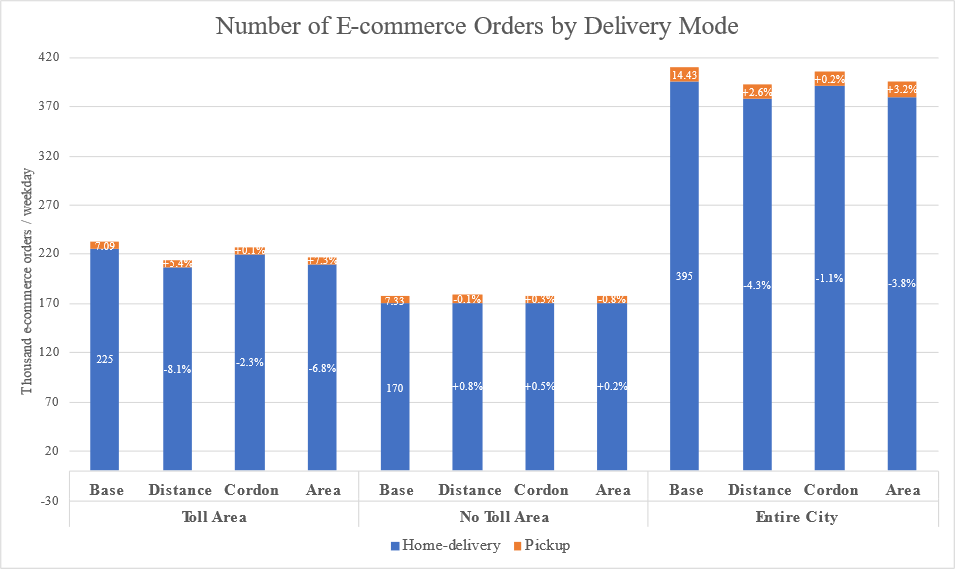}
  \caption{Impacts on E-commerce demand  \label{fig:EC_dem}}
\end{figure}

In terms of e-commerce, we find a decrease in total demand in the tolled area alongside with an even greater decrease of e-commerce home-delivery orders (Figure \ref{fig:EC_dem}). The total number of e-commerce orders decrease by 8.1\% under the distance-based scheme, 2.3\% under the cordon scheme, and 6.8\% under the area-based scheme. The main driver for this decrease is the increased delivery fees for home-delivery, which increases by 7.1\% under Distance, 1.8\% under Cordon, and 5.6\% under Area. The increase in pickup trips in counter-intuitive, since although pickup fees remain unchanged, the pickup trip is tolled which incurs extra costs. Overall, our results suggest that home-delivery becomes less preferred under the congestion pricing schemes for residents within the tolled area. 

  \begin{figure}[!h]
\hspace{-1 cm}
\includegraphics[scale=0.65]{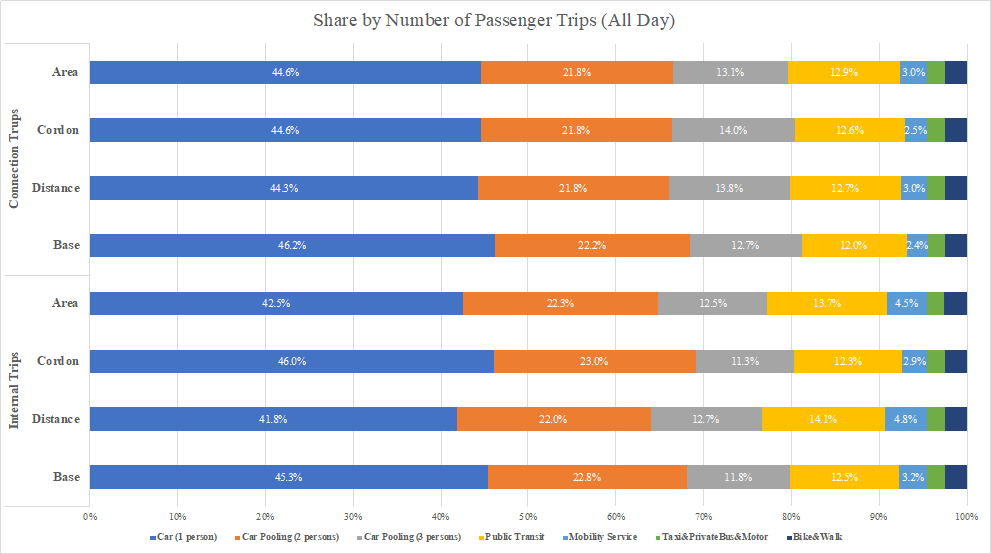}
  \caption{Impacts on mode shares  \label{fig:Mode_shares}}
\end{figure}

\subsubsection{Mode Shares}
The prototype city we study is highly auto-dependent (around 80\% in the Base scenario) but has a relatively high transit share (around 12\%) compared to most North American cities. The impacts of all three pricing schemes on mode shares is summarized in Figure \ref{fig:Mode_shares}. The distance-based and area-based schemes lead to the largest decreases in single occupancy trips for both internal trips (reduction in Car share by 3.5 and 2.8 share points, respectively) and  connection trips (1.9 and 1.6 share points, respectively). The decreases are the result of small shifts to public transit and relatively larger shifts to car-pooling. The cordon-based scheme is less effective in inducing mode shifts, once again underscoring the importance of cordon design. Moreover, the cordon scheme in fact results in a small increase in internal car trips due to the reduction in network congestion within the tolled region. The total number of car trips reduces by around 192,000 (6.6\%) under the distance-based scheme and 190,000 under the area-based scheme. 

\begin{table}[]
\renewcommand{\arraystretch}{0.75}
\small
\caption{Impacts on departure time}\label{table:DT_shifts}
\hspace{-0 cm}
\begin{tabular}{l|llll|llll}
\hline
\multirow{2}{*}{Mode} & \multicolumn{4}{c|}{Internal Trips}                                                                                                                                                                           & \multicolumn{4}{c}{Connection Trips}                                                                                                                                                                         \\
                      & Base   & Distance                                                        & Cordon                                                          & Area                                                            & Base   & Distance                                                        & Cordon                                                          & Area                                                            \\
                      \hline
\multicolumn{9}{c}{AM Peak}                                                                                                                                                                                                                                                                                                                                                                                                                         \\
\hline
Car                   & 16.4\% & \begin{tabular}[c]{@{}l@{}}15.7\%\\    \\ (-4.3\%)\end{tabular} & \begin{tabular}[c]{@{}l@{}}16.5\%\\    \\ (+0.8\%)\end{tabular} & \begin{tabular}[c]{@{}l@{}}16.6\%\\    \\ (+1.1\%)\end{tabular} & 16.7\% & \begin{tabular}[c]{@{}l@{}}16.2\%\\    \\ (-3.1\%)\end{tabular} & \begin{tabular}[c]{@{}l@{}}16.2\%\\    \\ (-3.1\%)\end{tabular} & \begin{tabular}[c]{@{}l@{}}16.8\%\\    \\ (+0.9\%)\end{tabular} \\
\hline
PT                    & 20.5\% & \begin{tabular}[c]{@{}l@{}}21.2\%\\    \\ (+3.1\%)\end{tabular} & \begin{tabular}[c]{@{}l@{}}20.3\%\\    \\ (-1.1\%)\end{tabular} & \begin{tabular}[c]{@{}l@{}}20.3\%\\    \\ (-1.2\%)\end{tabular} & 18.8\% & \begin{tabular}[c]{@{}l@{}}19.3\%\\    \\ (+2.7\%)\end{tabular} & \begin{tabular}[c]{@{}l@{}}19.1\%\\    \\ (+1.6\%)\end{tabular} & \begin{tabular}[c]{@{}l@{}}18.6\%\\    \\ (-0.9\%)\end{tabular} \\
\hline
Freight               & 10.1\% & \begin{tabular}[c]{@{}l@{}}10.0\%\\    \\ (-1.2\%)\end{tabular} & \begin{tabular}[c]{@{}l@{}}10.1\%\\    \\ (-0.0\%)\end{tabular} & \begin{tabular}[c]{@{}l@{}}10.5\%\\    \\ (+3.9\%)\end{tabular} & 11.1\% & \begin{tabular}[c]{@{}l@{}}11.0\%\\    \\ (-0.8\%)\end{tabular} & \begin{tabular}[c]{@{}l@{}}10.9\%\\    \\ (-1.8\%)\end{tabular} & \begin{tabular}[c]{@{}l@{}}11.4\%\\    \\ (+2.2\%)\end{tabular} \\
\hline
\multicolumn{9}{c}{PM Peak}                                                          \\
\hline
Car                   & 22.9\% & \begin{tabular}[c]{@{}l@{}}22.1\%\\    \\ (-3.3\%)\end{tabular} & \begin{tabular}[c]{@{}l@{}}23.0\%\\    \\ (+0.4\%)\end{tabular} & \begin{tabular}[c]{@{}l@{}}23.1\%\\    \\ (+0.9\%)\end{tabular} & 23.8\% & \begin{tabular}[c]{@{}l@{}}23.1\%\\    \\ (-2.8\%)\end{tabular} & \begin{tabular}[c]{@{}l@{}}23.5\%\\    \\ (-1.1\%)\end{tabular} & \begin{tabular}[c]{@{}l@{}}24.0\%\\    \\ (+0.7\%)\end{tabular} \\
\hline
PT                    & 27.1\% & \begin{tabular}[c]{@{}l@{}}27.6\%\\    \\ (+1.8\%)\end{tabular} & \begin{tabular}[c]{@{}l@{}}26.8\%\\    \\ (-0.9\%)\end{tabular} & \begin{tabular}[c]{@{}l@{}}26.9\%\\    \\ (-0.6\%)\end{tabular} & 24.3\% & \begin{tabular}[c]{@{}l@{}}24.8\%\\    \\ (+1.8\%)\end{tabular} & \begin{tabular}[c]{@{}l@{}}24.4\%\\    \\ (+0.1\%)\end{tabular} & \begin{tabular}[c]{@{}l@{}}24.2\%\\    \\ (-0.8\%)\end{tabular} \\
\hline
Freight               & 44.0\% & \begin{tabular}[c]{@{}l@{}}43.3\%\\    \\ (-1.6\%)\end{tabular} & \begin{tabular}[c]{@{}l@{}}43.9\%\\    \\ (-0.4\%)\end{tabular} & \begin{tabular}[c]{@{}l@{}}45.1\%\\    \\ (+2.4\%)\end{tabular} & 46.4\% & \begin{tabular}[c]{@{}l@{}}45.9\%\\    \\ (-1.0\%)\end{tabular} & \begin{tabular}[c]{@{}l@{}}45.4\%\\    \\ (-2.2\%)\end{tabular} & \begin{tabular}[c]{@{}l@{}}46.9\%\\    \\ (+1.0\%)\end{tabular} \\
\hline
\end{tabular} 

\end{table}
\subsubsection{Departure-time Patterns}
Table \ref{table:DT_shifts} summarizes the impacts of the three pricing schemes on departure time patterns for both passenger and freight traffic. In line with findings on overall welfare impacts and reduction in VKT, the distance-based scheme yields the largest reductions in the share of peak period travel (4.3\% and 3.1\% for internal and connection trips, respectively). The cordon-based scheme is equally effective in reducing the share of peak-period connection trips (3.1\% decrease) but in fact causes a small increase in the percentage of peak-period internal trips due to the reduction in travel times within the tolled region. Interestingly, the area-based scheme is ineffective in causing a reduction in the share of peak-period travel due to the flat toll rate throughout the day, and in fact, leads to an increase in the share of both connection and internal peak-period trips. Thus, although overall VKT does reduce under the area-based scheme, a large part of this reduction comes from the off-peak period. 

The departure time pattern of freight traffic is different from that of passengers -- the proportion of AM peak trips is quite low ($\sim$16\%) and most trips take place during mid-day and the PM peak ($\sim$44\%). Under congestion pricing, freight trips departing in AM and PM peaks decrease under Distance but the relative magnitudes are smaller than that of passenger trips, due to lower cost elasticities. The percentage decrease is larger in the PM peak where most of the traffic occurs. Under Cordon and Area, the impacts are similar to that of passenger trips but with smaller magnitudes of change. 

\begin{table}[]
\renewcommand{\arraystretch}{0.85}
\small
\caption{Shipment Size}\label{table:Ship_size}
\hspace{-0.5 cm}
\begin{tabular}{l|llll|llll}
\hline
\multirow{2}{*}{Type} & \multicolumn{4}{c|}{Internal Shipments}                                                                                                                                                                      & \multicolumn{4}{c}{Connection Shipments}                                                                                                                                                                    \\

                          & Base  & Distance                                                        & Cordon                                                          & Area                                                            & Base  & Distance                                                        & Cordon                                                          & Area                                                            \\
                          \hline
All                       & 149.3 & \begin{tabular}[c]{@{}l@{}}153.9 \\    \\ (+3.1\%)\end{tabular} & \begin{tabular}[c]{@{}l@{}}151.0 \\    \\ (+1.1\%)\end{tabular} & \begin{tabular}[c]{@{}l@{}}154.2 \\    \\ (+3.3\%)\end{tabular} & 154.3 & \begin{tabular}[c]{@{}l@{}}156.4 \\    \\ (+1.4\%)\end{tabular} & \begin{tabular}[c]{@{}l@{}}156.3 \\    \\ (+1.3\%)\end{tabular} & \begin{tabular}[c]{@{}l@{}}158.8 \\    \\ (+2.9\%)\end{tabular} \\
\hline
Non-ecommerce   shipments & 287.2 & \begin{tabular}[c]{@{}l@{}}293.3 \\    \\ (+2.1\%)\end{tabular} & \begin{tabular}[c]{@{}l@{}}290.8 \\    \\ (+1.2\%)\end{tabular} & \begin{tabular}[c]{@{}l@{}}295.5 \\    \\ (+2.9\%)\end{tabular} & 312.5 & \begin{tabular}[c]{@{}l@{}}318.3 \\    \\ (+1.9\%)\end{tabular} & \begin{tabular}[c]{@{}l@{}}317.8 \\    \\ (+1.7\%)\end{tabular} & \begin{tabular}[c]{@{}l@{}}320.4 \\    \\ (+2.5\%)\end{tabular} \\
\hline
E-commerce   shipments    & 14.1  & \begin{tabular}[c]{@{}l@{}}14.7 \\    \\ (+4.3\%)\end{tabular}  & \begin{tabular}[c]{@{}l@{}}14.5 \\    \\ (+2.9\%)\end{tabular}  & \begin{tabular}[c]{@{}l@{}}14.8 \\    \\ (+4.7\%)\end{tabular}  & 15.6  & \begin{tabular}[c]{@{}l@{}}15.9 \\    \\ (+1.9\%)\end{tabular}  & \begin{tabular}[c]{@{}l@{}}15.9 \\    \\ (+1.9\%)\end{tabular}  & \begin{tabular}[c]{@{}l@{}}16.1 \\    \\ (+3.2\%)\end{tabular} \\
\hline
\end{tabular}
\end{table}

\subsection{Logistics Operations}
In this section, we examine the impacts of the pricing schemes on logistics operations including shipment sizes and freight vehicle load factors. 

\subsubsection{Shipment Size}
Table \ref{table:Ship_size} shows the average shipment size by weight (kg) for different shipment types (B2B non-ecommerce or B2C e-commerce shipments) and shipment OD pairs. All the three pricing schemes result in an increase in shipment weight, although by differing magnitudes. Since the total commodity production and consumption is assumed to be fixed, a larger shipment size implies a lower shipment frequency and hence, a decrease in freight vehicle trips. For e-commerce shipments, which are significantly smaller than non-ecommerce shipments in size, this relative size increase is more significant. The increase is the most significant under the area-based and distance-based schemes whereas the changes are relatively small under the cordon-based scheme. 

\subsubsection{Freight Vehicle Load Factor}
The average freight vehicle load factors across all tours by vehicle type (LGV, HGV, and VHGV) are summarized in Table \ref{table:Freight_LF}. The OD pairs are the ODs of the first trip of the vehicle’s tour. Overall, the freight vehicle load factors are quite low for all vehicle types ($\sim$50\% for LGV and HGV, 32\%-36\% for VHGV) in the base scenario, indicating low operational efficiency. The load factors increase significantly only under Distance , which results in an increase in load factors of internal and connection tours by 2\%-6\%.

In the VOP model, we consider 3 constraints in shipment-to-vehicle assignment and tour formation: shipment time window, vehicle capacity, and vehicle maximum working hours. In the case of tour termination, we find a majority are due to the shipment time window constraint (26.3\%) and having no more shipments left (57.9\%). This might partly explain the reason carriers do not further consolidate shipments to one vehicle. We then select a subset of “large carriers” who have sufficiently large fleets and large numbers of shipments to be carried ($>$100 shipments). For these carriers, the change in vehicle load factor is much more significant (Table  \ref{table:Freight_LF}) compared to the entire carrier population under congestion pricing. Among the three freight vehicle types, in most cases the percentage increase in load factor is $VHGV > HGV > LGV$, indicating that for the $VHGV$ vehicle type with the highest toll rate, the load needs to be sufficiently improved than the status quo for the carriers to opt to use it.

\begin{table}[!h]
\renewcommand{\arraystretch}{0.85}
\small
\caption{Average freight Vehicle Load Factor}\label{table:Freight_LF}
\hspace{-0 cm}

\begin{tabular}{lllllllll}
\hline
\multirow{2}{*}{} & \multicolumn{4}{c}{Internal Tours}                                                                                                                                                                               & \multicolumn{4}{c}{Connection Tours}                                                                                                                                                                           \\
                  & Base   & Distance                                                          & Cordon                                                           & Area                                                             & Base   & Distance                                                         & Cordon                                                          & Area                                                             \\
                  \hline
\multicolumn{9}{c}{All carriers}                                                                                          \\
\hline
LGV               & 51.2\% & \begin{tabular}[c]{@{}l@{}}52.7\% \\    \\ (+2.8\%)\end{tabular}  & \begin{tabular}[c]{@{}l@{}}51.0\% \\    \\ (-0.4\%)\end{tabular} & \begin{tabular}[c]{@{}l@{}}51.1\% \\    \\ (-0.2\%)\end{tabular} & 53.0\% & \begin{tabular}[c]{@{}l@{}}54.0\% \\    \\ (+1.9\%)\end{tabular} & \begin{tabular}[c]{@{}l@{}}53.5\%\\    \\ (+1.0\%)\end{tabular} & \begin{tabular}[c]{@{}l@{}}53.2\%\\    \\ (+0.3\%)\end{tabular}  \\
HGV               & 49.3\% & \begin{tabular}[c]{@{}l@{}}51.7\% \\    \\ (+4.9\%)\end{tabular}  & \begin{tabular}[c]{@{}l@{}}49.0\% \\    \\ (-0.6\%)\end{tabular} & \begin{tabular}[c]{@{}l@{}}49.0\% \\    \\ (-0.6\%)\end{tabular} & 48.7\% & \begin{tabular}[c]{@{}l@{}}49.7\% \\    \\ (+2.0\%)\end{tabular} & \begin{tabular}[c]{@{}l@{}}49.4\%\\    \\ (+1.4\%)\end{tabular} & \begin{tabular}[c]{@{}l@{}}49.0\% \\    \\ (+0.7\%)\end{tabular} \\
VHGV              & 31.6\% & \begin{tabular}[c]{@{}l@{}}33.4\% \\    \\ (+5.9\%)\end{tabular}  & \begin{tabular}[c]{@{}l@{}}31.5\% \\    \\ (-0.3\%)\end{tabular} & \begin{tabular}[c]{@{}l@{}}31.1\% \\    \\ (-1.6\%)\end{tabular} & 35.8\% & \begin{tabular}[c]{@{}l@{}}36.7\%\\    \\ (+2.5\%)\end{tabular}  & \begin{tabular}[c]{@{}l@{}}36.6\%\\    \\ (+2.1\%)\end{tabular} & \begin{tabular}[c]{@{}l@{}}36.1\%\\    \\ (+0.9\%)\end{tabular}  \\
\hline
\multicolumn{9}{c}{Large   carriers}                                                                                      \\
\hline
LGV               & 71.7\% & \begin{tabular}[c]{@{}l@{}}75.9\% \\    \\ (+5.8\%)\end{tabular}  & \begin{tabular}[c]{@{}l@{}}70.9\%\\    \\ (-1.2\%)\end{tabular}  & \begin{tabular}[c]{@{}l@{}}71.3\%\\    \\ (-0.6\%)\end{tabular}  & 75.3\% & \begin{tabular}[c]{@{}l@{}}78.7\%\\    \\ (+4.4\%)\end{tabular}  & \begin{tabular}[c]{@{}l@{}}77.3\%\\    \\ (+2.7\%)\end{tabular} & \begin{tabular}[c]{@{}l@{}}76.2\%\\    \\ (+1.1\%)\end{tabular}  \\
HGV               & 64.9\% & \begin{tabular}[c]{@{}l@{}}69.4\% \\    \\ (+6.8\%)\end{tabular}  & \begin{tabular}[c]{@{}l@{}}63.9\%\\    \\ (-1.6\%)\end{tabular}  & \begin{tabular}[c]{@{}l@{}}64.4\%\\    \\ (-0.9\%)\end{tabular}  & 65.1\% & \begin{tabular}[c]{@{}l@{}}69.2\%\\    \\ (+6.4\%)\end{tabular}  & \begin{tabular}[c]{@{}l@{}}67.4\%\\    \\ (+3.6\%)\end{tabular} & \begin{tabular}[c]{@{}l@{}}67.0\%\\    \\ (+3.0\%)\end{tabular}  \\
VHGV              & 46.3\% & \begin{tabular}[c]{@{}l@{}}51.7\% \\    \\ (+11.7\%)\end{tabular} & \begin{tabular}[c]{@{}l@{}}45.4\% \\    \\ (-1.8\%)\end{tabular} & \begin{tabular}[c]{@{}l@{}}46.0\%\\    \\ (-0.6\%)\end{tabular}  & 54.3\% & \begin{tabular}[c]{@{}l@{}}58.2\%\\    \\ (+7.2\%)\end{tabular}  & \begin{tabular}[c]{@{}l@{}}57.0\%\\    \\ (+5.1\%)\end{tabular} & \begin{tabular}[c]{@{}l@{}}56.0\%\\    \\ (+3.3\%)\end{tabular} \\
\hline
\end{tabular}
\end{table}

\section{Conclusions}\label{sec:X}
This paper examines the impacts of several third-best congestion pricing schemes (distance-based, area-based and cordon) on both passenger transport and freight in an integrated manner. We employ the agent- and activity-based microsimulator SimMobility, which explicitly simulates the behavioral decisions of the entire population of individuals and business establishments, dynamic multimodal network performance, and their interactions.  

Simulations of a prototypical North American city with 4.6 million individuals and 0.130 million business establishments indicate that the distance-based scheme yields higher welfare gains (around 2.27 million USD per weekday) compared to the area-based (1.72 million USD per weekday) and cordon scheme (0.167 million USD). In line with several prior studies we find that the overall welfare gains are a modest fraction of toll revenues (30\%, 24\% and 11\% for the distance-based, area-based and cordon schemes, respectively). All three schemes are regressive with regard to both individuals and shippers, although the area-based scheme fares relatively better than the distance-based and cordon-schemes in terms of distributional impacts. The distance- and area-based schemes yield significant reductions in both VKT (10\% and 5\%, respectively for internal trips and 5.2\% and 2.6\%, respectively for connection trips) and peak-period travel time indices (by up to 13.1\% and 8.5\%, respectively).

Overall, given that the overall welfare gains are dwarfed by the toll revenues, we concur with \cite{eliasson2006equity} and \cite{santos2004distributional} who point out that ultimately, the outcomes from pricing depend in large part on how the toll revenues are used. In this regard and given our findings on the regressivity of all three schemes, a critical area for future research is in the design of suitable revenue recycling schemes that guarantee Pareto improvement. This would to a large degree placate public and political opposition. Tradable credit schemes also hold promise in this context since in principle, distributional outcomes and equity impacts can be appropriately influenced by the initial allocation of credits \citep{de2018congestion,seshadri2022congestion,CHEN2023104121}. Finally, it should be pointed out that our analysis has looked solely at the transportation system. \cite{de2011traffic} note that labor markets are distorted by income taxes, which have implications for pricing of commuting trips, and these effects may have first-order importance.

\section*{Acknowledgements}
This research was funded by Cintra S. A. We gratefully acknowledge their support. 

\bibliography{main}

\section*{Appendix A}\label{sec:App_A}
\begin{equation}
\underset{x_{ij}, \forall i,j}{\textnormal{min}}  \;\;  \sum_{i,j} x_{ij} 
\end{equation}
s.t
\begin{equation*}
\sum_{j} x_{ij} \geq  \underset{k}{\textnormal{max}} \;\; a_{ik}  \forall i
\end{equation*}
\begin{equation*}
\sum_{i} x_{ij} \geq  \underset{k}{\textnormal{max}} \;\; b_{jk}  \forall j
\end{equation*}
\begin{equation*}
 x_{ij} \in [0,1]  \;\; \forall i, j
\end{equation*}

where 
$x_{ij}$: binary variable, 1 if mapping between industry type $i$ and property type $j$ is feasible, 0 otherwise \\
$a_{ik}$: 	1 if industry type $i$ exists in zip code $k$, 0 otherwise \\
$b_{jk}$: 	1 if property type $j$ exists in zip code $k$, 0 otherwise \\

Note that additional constraints based on judgement are required, for example, a restaurant (industry type) cannot exist in a factory (property type). 

\section*{Appendix B}\label{sec:App_B}
\begin{equation}
\textnormal{min} \sum_{j,k} \left( \sum_i f_{ik}q_{ijk} x_{ij} - f^B_{jk}   \right)^2 
\end{equation}
s.t
\begin{equation*}
 r_{ik} = c_i + d_k, \;\; \forall i, k
\end{equation*}
\begin{equation*}
 f_{ik} =  r_{ik}N^E_{ik} + p_{ik}N^T_{ik}, \;\; \forall i, k
\end{equation*}
\begin{equation*}
\sum_{j} q_{ijk}x_{ij} = 1 \;\;  \forall i, k
\end{equation*}
\begin{equation*}
q_{ijk} \leq x_{ij} \;\; \forall i, j, k
\end{equation*}
\begin{equation*}
 q_{ijk} \in [0,1]  \;\; \forall i, j, k
\end{equation*}
\begin{equation*}
 c_i \geq 0, d_k \geq 0, p_{ik} \geq 0,  \;\; \forall i, k
\end{equation*} 
Decision variables \\
$r_{ik}$: conversion rate for industry $i$ in zip code $k$ \\
$c_i$: part of rate  $r_{ik}$ specific to industry $i$ \\
$d_k$: part of rate  specific to zip $k$ \\
$p_{ik}$: constant of floor conversion rate per establishment for industry $i$ in zip code $k$  \\
$f_{ik}$: establishment floor area of industry $i$ in zip code $k$  \\
$q_{ijk}$: proportion of floor area of industry $i$ assigned to property type $j$ in zip code $k$  \\
Inputs: \\
$N^E_{ik}$: number of employees in industry $i$ in zip code $k$  \\
$N^T_{ik}$: number of establishments of industry $i$ in zip code $k$ \\
$f^B_{jk}$: building floor area of property $j$ in zip code $k$ \\
$x_{ij}$: estimated mapping relationship -- 1 if mapping between industry type $i$ and property type $j$ is feasible, 0 otherwise \\

Following this, the estimated floor area of an establishment $m$ of industry type $i$ in zip code $k$, denoted by  $\hat{f}_{ikm}$ is given by $\hat{f}_{ikm} = r_{ik}N^E_{ikm} + p_{ik} $.


\section*{Appendix C}\label{sec:App_C2}
\begin{equation}
\textnormal{min} \sum_{n} \left( \sum_m y_{mn}\hat{f}_{km}  - f^B_{kn}   \right)^2 
\end{equation}
s.t
\begin{equation*}
\sum_{n} y_{mn} = 1 \;\;  \forall m
\end{equation*}
\begin{equation*}
y_{mn} \leq x_{i(m)j(n)} \;\; \forall m, n
\end{equation*}
\begin{equation*}
 y_{mn} \in [0,1]  \;\; \forall i, j, k
\end{equation*}
Decision variables \\
$y_{mn}$: binary variable, 1 if establishment $m$ is in building $n$, 0 otherwise \\
Inputs: \\
$f^B_{kn}$:  floor area of building $n$ in zip code $k$ \\
$\tilde{f}^{'}_{km}$: adjusted floor area of establishment $m$ in zip code $k$ \\
$x_{i(m)j(n)}$: 1 if establishment $m$ of industry type $i$ can be mapped to buidling $n$ of property type $j$, 0 otherwise \\

\end{document}